\documentclass[]{pasj01}
\usepackage{array}

\Received{2023/07/17}
\Accepted{2023/08/30}
\Published{2023/mm/dd}
 
\begin{document} 

\title{ 
FLASHING: Project Overview
}

\author{
Hiroshi \textsc{Imai}\altaffilmark{1,2}, 
Yuhki \textsc{Hamae}\altaffilmark{3}, 
Kei \textsc{Amada}\altaffilmark{4}, 
Keisuke \textsc{Nakashima}\altaffilmark{4},
Ka-Yiu \textsc{Shum}\altaffilmark{4}, 
Rina \textsc{Kasai}\altaffilmark{3}, 
Jos\'e F.\ \textsc{G\'omez}\altaffilmark{5}, 
Lucero \textsc{Uscanga}\altaffilmark{6}, 
Daniel \textsc{Tafoya}\altaffilmark{7}, 
Gabor \textsc{Orosz}\altaffilmark{8}, 
and Ross A. \textsc{Burns}\altaffilmark{9,10}
}

\altaffiltext{1}{Amanogawa Galaxy Astronomy Research Center, Graduate School of Science and Engineering, Kagoshima University, 
1-21-35 Korimoto, Kagoshima 890-0065}
\email{hiroimai@km.kagoshima-u.ac.jp}

\altaffiltext{2}{Center for General Education, Institute for Comprehensive Education, Kagoshima University,  
1-21-30 Korimoto, Kagoshima 890-0065}

\altaffiltext{3}{Faculty of Science, Kagoshima University, 
1-21-35 Korimoto, Kagoshima 890-0065}

\altaffiltext{4}{Graduate School of Science and Engineering, Kagoshima University, 
1-21-35 Korimoto, Kagoshima 890-0065}

\altaffiltext{5}{Instituto de Astrof\'isica de Andaluc\'{\i}a, CSIC, Glorieta de la Astronom\'{\i}a s/n, E-18008 Granada, Spain}

\altaffiltext{6}{Departamento de Astronom\'ia, Universidad de Guanajuato, A.P. 144, 36000 Guanajuato, Gto., Mexico}

\altaffiltext{7}{Department of Space, Earth and Environment, Chalmers University of Technology, Onsala Space Observatory, 439 92 Onsala, Sweden}

\altaffiltext{8}
{Joint Institute for VLBI ERIC, Oude Hoogeveensedijk 4, 7991 PD Dwingeloo, The Netherlands}

\altaffiltext{9}
{RIKEN Cluster for Pioneering Research, 2-1, Hirosawa, Wako-shi, Saitama 351-0198, Japan}

\altaffiltext{10}{Department of Science, National Astronomical Observatory of Japan, 2-21-1 Osawa, Mitaka, Tokyo, 181-8588, Japan}


\maketitle

\begin{abstract}
This paper describes the overview of the FLASHING (Finest Legacy Acquisitions of SiO-/ H$_2$O-maser Ignitions by the Nobeyama Generation) project promoted using the 45~m telescope of Nobeyama Radio Observatory, which aims to intensively monitor H$_2$O (22~GHz) and SiO (43~GHz) masers associated with so-called ``water fountain" sources. Here we show scientific results on the basis of the data taken in for the first five  seasons of FLASHING, from 2018 December to 2023 April). We have found the evolution of the H$_2$O maser spectra, such as new spectral components breaking the record of the jet's top speed and/or systematic velocity drifts in the spectrum indicating acceleration or deceleration of the maser gas clumps. For the 43~GHz SiO maser emission, we have found its new detection in a source while its permanent disappearance in other source. Our finding may imply that the jets from these water fountains can be accelerated or decelerated, and show how cicumstellar envelopes have been destroyed.
\end{abstract}

\section{Introduction}
\label{sec:introduction}
~~~The ``water fountain" (WF) sources are 22~GHz H$_2$O maser sources associated with bipolar fast jets in the transition from the final phase of an asymptotic giant branch (AGB) star to the early phase of a planetary nebula (PN). These jets are faster expanding than the circumstellar envelopes (CSEs) traced in 1612~MHz OH masers at a typical velocity of 10--30~km~s$^{-1}$ (e.g., \cite{ima07a}). So far only 15 WFs have been confirmed \citep{gom17}, implying that the duration in which they are visible as high velocity circumstellar H$_2$O masers is extremely short ($<$100~yr, \cite{kho21}). It has been recently proposed that a large fraction of binary systems in the Milky Way that create common envelopes can become the WFs at some point in their evolution \citep{kho21}. Some of these WFs also show H$_2$O maser emission at millimeter/submillimeter wavelengths \citep{taf14}. 

Interferometric observations of these H$_2$O masers have found a wide variety of substructures of the maser spatio-kinematics such as multiple arcs, pairs of maser feature groups distributed point symmetrically, groups of maser features containing a large internal velocity dispersion ($\gg$10~km~s$^{-1}$, e.g., \cite{ima02a,cla09,ima13b,oro19,ima20}, hereafter Paper I). Some of the WFs, W~43A and IRAS~16552$-$3050, also host 43~GHz SiO ($v=1$ $J=1\rightarrow 0$) maser emission (\cite{nak03,ima05,ama22}, hereafter Paper II). They imply  different types of the central stellar systems driving their hosting jets and/or different stages of the jet evolution. 

Recent observations of thermal molecular line and dust emission with the Atacama Large Millimeter-and-submillimeter Array (ALMA) have directly revealed that H$_2$O masers in the WFs are associated with the interaction of the fast jet with the parental CSE sculptured by the jet \citep{sah17,taf19,taf20}. However, the origin of H$_2$O maser excitation is still obscure in detail because their associated regions exhibit deceleration in some sources and acceleration in others (e.g., \cite{gom15, oro19}).
 
Taking into account the rarity of the WFs and the possibility of emergence of new H$_2$O maser sources and spectral evolutions in the known WFs on the short timescale mentioned above, monitoring of these H$_2$O masers is crucial for understanding the phenomena of the WFs in a unified scheme modeled with the information of jet development and/or devolution. In fact, in W~43A, the growth rate of the jet is reliably determined through the combination of the maser proper motions and the change rate of a total length of the maser jet on the sky \citep{ima07b,cho15,taf20}. A weekly--biweekly monitoring program of the WF H$_2$O and SiO masers over decades is ideal to trace the evolution of the WF systems, but it is still too expensive in interferometric observations.  

Therefore, we have conducted a project so-called FLASHING (Finest Legacy Acquisitions of SiO-/H$_2$O-maser Ignitions by Nobeyama Generation) with the 45~m telescope of Nobeyama Radio Observatory (NRO), aiming at single-dish monitoring of the WF maser spectra (Paper I). This has been launched soon after the new quasi-optics has been commissioned, enabling us to simultaneously observe in the 22~GHz and 43~GHz bands \citep{oka20,ima22}. Although most of the WFs host only H$_2$O masers, monitoring of SiO maser emission is also interesting to understand their property in W~43A and discover new surprising maser ignitions as already reported (Paper II). The weakness of some velocity components of H$_2$O masers as well as SiO masers in the WFs requires a high sensitivity telescope such as the 45~m telescope. This project, and future ones using other telescopes, will be continued until either the telescope or our team being fully decommissioned. 

This paper describes the overview of FLASHING and presents its early scientific results based on the spectral data taken during 2018 December--2023 April. Sect.\ \ref{sec:observations} describes the FLASHING observations in detail. Sect.\ \ref{sec:results} shows the results. Sect.\  \ref{sec:discussion} discusses the results and presents the perspectives of FLASHING and related projects. 

\section{Observations and data reduction}
\label{sec:observations}

FLASHING is a monitoring observation project targeting different evolutionary stages of evolved stars showing circumstellar H$_2$O and SiO maser emission, with a special focus on the WFs. These masers have been simultaneously observed in the K- (22~GHz, H$_2$O masers) and Q- (43~GHz, SiO masers) bands using the new quasi-optics system at the NRO 45~m telescope, mentioned above. In this paper, we report the results of the FLASHING observations from 2018 December to 2023 April. 

\begin{table}
  \caption{Maser sources monitored in FLASHING}
  \label{tab:sources}
  \begin{tabular}{c@{}c@{ }c@{ }c@{ }l@{ }r@{ }c@{ }c@{ }c@{}}
      \hline
       & & \multicolumn{6}{c}{Coordinates (J2000)} & \\
       ID\footnotemark[$\ast$]  &   Source name &  h &  m &      s &   $^{\circ}$ &  $^{\prime}$ & $^{\prime\prime}$ & Type\footnotemark[$\dagger$] \\ \hline
       \multicolumn{8}{c}{Water fountain sources} \\ \hline
        a & S CrB  &  15 & 21 & 23.95 & $+$31 & 22 & 02.5 & Mira \\
        b & IRAS16552$-$3050 & 16 & 58 &  27.80 & $-$30 & 55 &  06.2 & WF \\
        c & IRAS17021$-$3109 & 17 & 05 & 23.71 & $-$31 & 13 & 18.7 & OH/IR \\
        d & IRAS17291$-$2147 & 17 & 32 & 10.10 & $-$21 & 49 & 59.0 & WF \\
        e & IRAS17348$-$2906 & 17 & 38 & 03.92 & $-$29 & 08 & 16.5 & OH/IR \\
        f & IRAS18043$-$2116 & 18 & 07 &  21.10 & $-$21 & 16 & 14.2 & WF\\
        g & IRAS18056$-$1514 & 18 & 08 & 28.39 & $-$15 & 13 & 28.3 & WFC \\
        h & IRAS18113$-$2503 & 18 & 14 & 27.26 & $-$25 & 03 & 00.4 & WF\\
        i &     OH12.8$-$0.9 & 18 & 16 & 49.23 & $-$18 & 15 & 01.8 & WF\\
        j &  IRAS18251$-$1048 & 18 & 27 &   56.30 & $-$10 & 46 &  58.0 & OH/IR \\
        k & IRAS18286$-$0959 & 18 & 31 & 22.93 &  $-$09 & 57 & 21.7 & WF \\
        l &    OH 16.3$-$3.0 & 18 & 31 & 31.49 & $-$16 &  08 & 46.23 & WFC \\ 
        m &             W43A & 18 & 47 & 41.16 &  $-$01 & 45 & 11.6 & WF \\
        n & IRAS18455$+$0448 & 18 & 48 & 02.30 & $+$04 & 51 & 30.5 & WF \\
        o & IRAS18460$-$0151 & 18 & 48 &  42.80 &  $-$01 & 48 & 40.0 & WF \\
        p & IRAS18596$+$0315 & 19 & 02 &  06.28 &  $+$03 & 20 & 16.3 & WF \\
        q &  IRAS19067$+$0811 & 19 & 09 &  07.47 & $+$08 & 16 &  22.5 & WFC \\
        r & IRAS19134$+$2131 & 19 & 15 & 35.22 &  $+$21 & 36 & 33.9 &WF \\
        s & IRAS19190$+$1102 & 19 & 21 & 25.09 &  $+$11 & 08 & 41.0 &WF \\
        t &          K3$-$35 & 19 & 27 & 44.02 & $+$21 & 30 & 03.4 & PN \\
        u &  IRAS19312$+$1950 & 19 & 33 &  24.25 & $+$19 & 56 &  55.7 & PAGB\footnotemark[$^{\#}$] \\
        v & IRAS19356$+$0754 & 19 & 38 & 01.85 & $+$08 & 01 & 32.8 & WFC \\
    \hline
    \end{tabular}

       \noindent
\footnotemark[$\ast$]Source ID label referenced in Table \ref{tab:obslogs}. \\
\footnotemark[$\dagger$]Source type: WF: Water fountain source;  WFC: Water fountain candidate; OH/IR: OH/IR star; PAGB: Post-AGB star; PN: Planetary nebula. \\
\footnotemark[$^{\#}$]There exists a controversy about the source type. See \citet{qiu23}.\\
\end{table}

A total of 22 objects have been monitored. This sample included the 12 WFs observable from NRO (declination $\geq -30$\arcdeg). The rest of the targets were four candidates to being WFs, in which the H$_2$O maser spectrum had a velocity width larger than that of 1612~MHz OH masers \citep{yun13,vle14} but for which there still exists no reported interferometric observation to confirm that the H$_2$O masers trace a jet, and other objects such as OH/IR stars, and post-AGB stars (including PNs hosting H$_2$O masers). These additional sources were mainly observed to fill gaps between allocated observation time slots of ours and other projects and when no WF was visible. Table \ref{tab:sources} gives the list of these observed sources. 

For H$_2$O masers in the K-band, right- and left-hand circular polarization was observed, and the corresponding spectra were averaged to improve sensitivity. On the other hand, only left-hand circular polarization spectra were obtained in the Q-band observations of SiO masers, when observed with the H40 receiver (before 2022 March and in some observations before September 2022). In the rest of the Q-band observations, SiO masers have been observed with the Z45 receiver \citep{nak15}, which can receive in two orthogonal linear polarization signals and yields higher sensitivity (Paper II; \cite{ima22}). The received signals were digitized using the PANDA sampler (OCTAD) and processed using 16 arrays of the SAM45 digital spectrometer. Thus we could obtain simultaneously several spectra of molecular lines.  Table \ref{tab:antSpec} gives the beam size and the aperture efficiency of the telescope, and the bandwidths and velocity channel resolutions of the spectroscopy. Table \ref{tab:maserLine} gives the list of the observed molecular lines. This paper describes only the results of the H$_2$O and SiO lines.  

The integration time of each observation toward the target source was 15--50~min, yielding root-mean-square (RMS) noise levels of 0.1--0.3~Jy and 0.2--0.6~Jy at the K- and Q-bands, respectively, for the maser spectra taken in a single observation block, and with the velocity resolutions of 0.41~km~s$^{-1}$ (K-band) and 0.41--0.43~km~s$^{-1}$ (Q-band). These sensitivity levels were chosen taking into account the available machine time to monitor all the WFs observable with the telescope (declination $\delta\geq -30\arcdeg$) and the faintest H$_2$O maser components (flux density $S_{\nu}\geq$0.2~Jy) that can be mapped with the existing high sensitivity VLBI networks. In fact, the median RMS noise level was 0.08--0.11~Jy in the H$_2$O maser spectra from season to season, 0.14--0.21~Jy and 0.08--0.11~Jy in the SiO maser spectra with the H40 and the Z45 receivers, respectively. 

\begin{table}
  \tbl{Observation specification and setup}{%
  \begin{tabular}{c@{ }c@{ }c@{ }c@{ }c@{ }c@{ }c@{}}
      \hline
    & $\nu_{\rm rx}$$^{*1}$ & $\Delta v_{\rm sp}$$^{*2}$  & $\Delta v$$^{*3}$     
    & $\theta_{\rm beam}$$^{*4}$  &  
    & $T_{\rm sys}$$^{*6}$ \\
Receiver  & (GHz)    & (km s$^{-1}$)   & (km s$^{-1}$) & (\arcsec)    
& $\eta_{\rm eff}$$^{*5}$ & (K) \\
\hline \hline
H22 & 20--24     & $\pm$820$^{*7}$  & 0.41  & 73  & 0.61$^{*8}$ & 100 \\
&& $\pm$420  &&&& 70$^{*10}$ \\
H40 & 42.5--44.5 & $\pm$420    & 0.41--0.43   & 39   & 0.55$^{*8}$ & 150 \\
Z45 & 40--46 & $\pm$420       & 0.41--0.43    & 37  & 0.50$^{*9}$ & 100 \\
\hline
    \end{tabular}}\label{tab:antSpec}
\begin{tabnote}
\footnotemark[$*1$] Frequency coverage of the receiver. \\
\footnotemark[$*2$] Velocity coverage of the spectral window. \\
\footnotemark[$*3$] Velocity {\bf resolution of} the spectrum {\bf after smoothing}. \\
\footnotemark[$*4$] The FWHM beam size. \\ 
\footnotemark[$*5$] Aperture efficiency. \\
\footnotemark[$*6$] Typical system noise temperature value at high elevations ($\geq$60\arcdeg).\\
\footnotemark[$*7$] With three spectrum windows available until 2022 April.  \\ 
\footnotemark[$*8$] Measured value in the 2018 season.\\
\footnotemark[$*9$] Measured value in the 2022 season in the H22$+$Z45 mode.\\
\footnotemark[$*10$] Value improved since 2022 September.
\end{tabnote}
\end{table}

\begin{table}
  \tbl{Observed spectral lines}{%
  \begin{tabular}{ll}
      \hline
            Molecule and transition & Center frequency$^{\dagger}$\\
            & (GHz) \\ \hline \hline
            H$_2$O $J_{K_-K_+} = 6_{12} \rightarrow 5_{23} $ & 22.235080 \\
            \multicolumn{1}{r}{(red shift side)$^{\S}$} & 22.205080 \\
            \multicolumn{1}{r}{(blue shift side)$^{\S}$} & 22.265080 \\
            NH$_3$ $(J,K)=(1,1)$ & 23.694 \\
            SiO $v = 3$ $J = 1 \rightarrow 0 $ & 42.519340 \\
            HC$_5$N $J=16 \rightarrow 15$$^{\S}$ & 42.602153 \\
            SiO $v = 2$ $J = 1 \rightarrow 0 $ 
            & 43.820539 \\
            $^{29}$SiO $v = 0$ $J = 1 \rightarrow 0 $ 
            & 42.879850 \\
            SiO $v = 1$ $J = 1 \rightarrow 0 $             
             & 43.122027 \\
            SiO $v = 0$ $J = 1 \rightarrow 0 $               & 43.423798 \\
            CCS $(N,J)=(4,3) \rightarrow (3,2)$$^{\S}$ & 43.981029 \\
            CH$_3$OH  $J_K = 7_{0}  \rightarrow 6_{1} A^{+}$ & 44.06941 \\
      \hline
    \end{tabular}}\label{tab:maserLine}
\noindent
$^{\dagger}$Center frequency of the spectral window to observe the listed line(s).\\
$^{\S}$Assigned until 2022 March when using the H22$+$H40 receiving mode.\\
\end{table}

In the first season (2018 December--2019 May), these observations were made as a Back-up Program project, namely under weather conditions unfavorable for projects at shorter millimeter bands, such as periods when strong winds caused large pointing offsets and/or the system had high noise temperatures. Although the time allocation was thus irregular and the observation cadence was also so irregular (observations every 1--8 months for each source), a total observation time of 260 hours were collected. In the subsequent seasons, after 2019 December, scheduled observations were conducted for 70--120 hours per year in a regular cadence (2--4 weeks). 

The spectral data were reduced using the JavaNewstar package\footnote
{https://www.nro.nao.ac.jp/\textasciitilde jnewstar/html/}. 
The raw data were at first split into the individual spectral windows and the spectra of each window were integrated along time. Then, fitting and subtraction of a line-emission-free baseline profile was performed with a seven-degree polynomial function. The conversion factors from the antenna temperature to the flux density unit were adopted to be 2.67~Jy~K$^{-1}$ and 3.20~Jy~K$^{-1}$ for the 22~GHz- and 43~GHz-band data, respectively (Paper I; Paper II). 

The calibrated spectra in the output text files of JavaNewstar were then automatically analyzed using our own program scripts, which found spectral channels whose flux density is higher than a 4-$\sigma$ level. In the case in which multiple spectral peaks may exist in the consecutive spectral channels with maser emission, local spectral peaks were identified as isolated or independent ones if each of them has a flux density higher by a 3-$\sigma$ level than that at the first adjacent local flux-density minimum traced from the local peak. Three consecutive channels were extracted as true maser detection even though only the center out of the three channels had a flux density higher than a 4-$\sigma$ level.

In order to more precisely determine the peak flux density and the corresponding line-of-sight velocity in each isolated spectral peak, data of three consecutive spectral channels were extracted around the channel with the local maximum flux density and used for fitting to a quadratic function, for simplicity without considering a spectral baseline level. 
Thus the flux density and the velocity at the spectral peak were derived from the fitting parameters. In this paper, we mainly track these individual peak flux densities and velocities. 
The velocity of a spectral peak determined in this simple fitting method may be affected by blending of multiple velocity components. However, taking into account the threshold of identification of the isolated peak mentioned above, the uncertainty of the peak velocity should be less than the velocity resolution. The relative uncertainty of the flux density scale may be $\leq$20\%, mainly due to  antenna pointing error ($\leq$1/5 of the antenna beam), while the uncertainty of the spectral peak velocity may be much better ($\leq$0.1~km~s$^{-1}$) than the spectral resolution. 
After getting the parameters of the spectral components mentioned above, we selected the components whose peak flux density exceeded a 5-$\sigma$ noise level for its firm identification. 

In Appendix \ref{sec:log}, the logs of the FLASHING observations are given to show the dates of the observations and the records of positive and negative detections of the masers. 

\section{Results}
\label{sec:results}

Although FLASHING is still ongoing and all of the results presented in this paper will be further inspected for more specific conclusions, we have obtained some interesting results on the properties of the masers in the WFs and the WF candidates. 

All of the 12 WFs have been detected at a few epochs at least in H$_2$O masers. Only one WF (IRAS~16552$-$3050) also has been detected in SiO masers. One WFC (IRAS~18056$-$1514) has been detected in H$_2$O masers, another (OH~16.3$-$3.0) in both H$_2$O and SiO masers. Two WFCs  (IRAS~19067$+$0811 and IRAS~19356$+$0754) have been detected in neither H$_2$O nor SiO masers. One OH/IR star (IRAS~17021$-$3109) has been detected in neither H$_2$O nor SiO masers, another OH/IR star (IRAS~17348$-$2906) in H$_2$O masers, and other OH/IR star (IRAS~18251$-$1048) in both H$_2$O and  SiO masers. 

The time series of the maser spectra are presented in Appendix \ref{sec:spectra}  and \ref{sec:spectra-SiO}.  

\begin{table}
  \caption{Fastest velocity components in H$_2$O maser spectra recorded in FLASHING}
  \label{tab:WF_velocities}
  \begin{tabular}{lrrrc} \hline \hline
        & $V_{\rm blue}$ &  $V_{\rm red}$ & $V_{\rm sys}$ & \\
WF name & \multicolumn{3}{c}{(km~s$^{-1}$)} & Ref.\footnotemark[2] \\ \hline
IRAS~16552$-$3050 &$  -78 $& 128 & 16 & j, k \\
IRAS~17291$-$2147 &$   -34 $&  36 & 21\footnotemark[3] & h \\
IRAS~18043$-$2116 &$ -145 $& 376 & 87 & l \\
IRAS~18113$-$2503 &$ -155 $& 343 & $\sim$90\footnotemark[5] & j \\
OH~12.8$-$0.9 & ...\footnotemark[1] & $-34$ & $-$56 & b \\
IRAS~18286$-$0959&$ -224$ &  345 & 19\footnotemark[4] & e, g \\
W~43A &$ -66  $& 134  & 34 & i \\
IRAS18455$+$0448 & 28 & 45 & 34 & m \\
IRAS18460$-$0151 &$ -64  $&  276 & 128 & d, f \\
IRAS18596$+$0315 & ...\footnotemark[1] &  115 & $\sim$90\footnotemark[5] & f, j \\
IRAS19134$+$2131 &$ -125  $&  $ -9 $  & $-$68\footnotemark[3] & c \\
IRAS19190$+$1102 &$ -15  $&  85 & 23 & a\\
\hline
\end{tabular}

\noindent
\footnotemark[1]No detection in FLASHING. \\
\footnotemark[2]Reference of the systemic velocity, 
a: \citet{lik89}; 
b: \citet{gom94}; 
c: \citet{ima07b}; 
d: \citet{ima13b};
e: \citet{ima13c}; 
f: \citet{riz13};
g: \citet{yun14}; 
h: \citet{gom15}; 
i: \citet{taf20};
j: \citet{kho21}; 
k: \citet{ama22}; 
l: \citet{usc23}; 
m: \citet{vle14}
\\
\footnotemark[3]Just adopting the center velocity between the most blue- and red-shifted components of H$_2$O masers. \\
\footnotemark[4]Adopting the center velocity between the two velocity components of 1612~MHz OH masers that were detected in different epochs.\\
\footnotemark[5]Eye inspection on the basis of the spectrum of the C$^{18}$O emission \citep{kho21}. 
\end{table}

\subsection{New high velocity components in H$_2$O maser spectra towards the WFs}
\label{sec:result-evolution}

Comparing with previous publications over decades, one can see the systematic evolution of H$_2$O  maser spectra in some WFs. The detection of new maser components at velocities higher than previously identified directly means to increase the estimated velocity for the stellar jet.

The top speed of a WF jet would be related to its powering mechanism and launching point. For instance, in the case of an accretion disk, it may be related to the location of the disk from which it is launched. Recently, thermal molecular line emissions from the WF jets have been directly mapped with ALMA in detail  (\cite{sah17,gom18,taf19,taf20,usc23}, hereafter Paper III). For the source in which the jet speed is estimated with thermal emission to be much higher than that indicated in H$_2$O maser spectrum, faster velocity components of the masers are expected when new blobs of gas are ejected in the jet. 

The detection of such new components may also indicate the rapid evolution of the jet over a few decades, given the short timescales of masers in the WFs ($<$100~yr). Actually, such new components are  very faint ($<$1~Jy) and short-lived (a month or shorter) at their beginnings. Such cases of new detections have already been published within the FLASHING project for IRAS~18043$-$2116 (Paper III) and IRAS~18286$-$0959 (Paper I). This paper presents the spectral evolution and detections of new maser components in other WFs.

Table \ref{tab:WF_velocities} gives the record of the most blue- and red-shifted components in the H$_2$O maser spectra obtained in FLASHING. $V_{\rm blue}$ and $V_{\rm red}$ give the local-standard-of-rest (LSR) velocities of the most blue- and red-shifted peaks of the H$_2$O maser spectra, respectively. For reference, the stellar systemic velocity estimated, $V_{\rm sys}$ is also given. 
As shown in table \ref{tab:antSpec}, the minimum total velocity width of 840~km~s$^{-1}$ is enough to cover the whole velocity range of each H$_2$O maser spectrum of the WFs. 

For IRAS~16552$-$3050, the previous observations of the H$_2$O masers \citep{sua08,sua09} found the maser emission around $V_{\rm LSR}=-70$-- $-40$~km~s$^{-1}$ and $V_{\rm LSR}=+$55-- $+$120~km~s$^{-1}$. Thus the components with the most extreme velocities in FLASHING break the record of the top speed of the outflow. 

For IRAS~17291$-$2147, \citet{gom15} and \citet{gom17} found the maser emission in two velocity groups between $V_{\rm LSR}=-28$-- $-19$~km~s$^{-1}$ and $V_{\rm LSR}=+$53-- 
$+$70~km~s$^{-1}$. Thus the blue-shifted components detected in FLASHING also break the record of the top speed of the outflow. 
 
For IRAS~18043$-$2116, the most blue- and red-shifted components have been detected at $V_{\rm LSR}\sim -165$~km~s$^{-1}$ and $\sim+$376~km~s$^{-1}$, respectively, as of 2021 March (Paper III). Thus the new data of FLASHING for the latest two years do not break the record of the top speed of the outflow. Note that the highest-velocity ($V_{\rm LSR}<-100$~km~s$^{-1}$ and $V_{\rm LSR}>+300$~km~s$^{-1}$), faint components had been invisible before the year 2015 while they are now growing. These blue-shifted components were confirmed at first by \citet{per17} in 2015 August and their flux densities have increased as seen in the FLASHING spectra (figure \ref{fig:IRAS18043-2116}). The red-shifted components have been confirmed in only FLASHING. 

For IRAS~18286$-$0959, the detection of the most red-shifted component has already been reported in Paper I, while the most blue-shifted component again breaks the record of the jet's top speed. At the present, this WF has the record of the total velocity spread of H$_2$O maser spectrum ($\Delta v\sim +$570~km~s$^{-1}$), overtaking  IRAS~18113$-$2503 ($\Delta v\sim $500~km~s$^{-1}$) and IRAS~18043$-$2116 ($\Delta v\sim +$540~km~s$^{-1}$). The spectral evolution of H$_2$O masers in IRAS~18286$-$0959 looks the most significant of all the WFs. The total velocity spread has doubled since the first discovery of this WF \citep{deg07}. Note that the total length of the spatial distribution of H$_2$O maser features also has been doubled (Paper I). 

For W~43A, the first observation by \citet{gen77} in 1977 found the most blue- and red-shifted components at $V_{\rm LSR}\sim -55$~km~s$^{-1}$ and $\sim+$130~km~s$^{-1}$, respectively. In the VLBA observations during 1994--2005, \citet{cho15} found the most blue- and red-shifted components at $V_{\rm LSR}\sim -62$~km~s$^{-1}$ and $\sim +$128~km~s$^{-1}$, respectively. The FLASHING spectra of the H$_2$O masers break the record of the jet's top speed. Here note that the total velocity spread of the maser components has increased by only $\sim +$15~km~s$^{-1}$ over 46~yr and finally gets comparable to that found in the CO emission $\Delta v\sim$200~km~s$^{-1}$ \citep{taf20}. 

For IRAS~18460$-$0151, the previous observations \citep{deg07,ima13c} found the maser emission in the range of $V_{\rm LSR}\sim-70$-- $+$245~km~s$^{-1}$ in three main velocity groups. The detection of the most red-shifted component in FLASHING breaks the record of the jet's top speed. 

For IRAS~19190$+$1102, the previous observations \citep{lik89,day10} found the maser emission around the range of $V_{\rm LSR}=-53$-- $+$78~km~s$^{-1}$. The detection of the most red-shifted component in FLASHING breaks the record of the jet's top speed. 

Contrary to the H$_2$O maser spectra of the WFs mentioned above, the velocity spread of the components in IRAS~18113$-$2503 looks stable since the early observations with the Green Bank Telescope (GBT) and the Very Large Array (VLA) (from $V_{\rm LSR}=-154$ to $+$351~km~s$^{-1}$, \cite{gom11,gom15}). This seems inconsistent with the interpretation by \citet{oro19} for the VLBA data that the H$_2$O masers are associated with gas clumps rapidly decelerated in the fast jet. The FLASHING observations have covered a total velocity range of 1640~km~s$^{-1}$ (table \ref{tab:antSpec}), but we could not find any new velocity components beyond the velocity range of the detected components mentioned above. 

For IRAS~19134$+$2131, the first observation by \citet{lik92} during 1988--1989 found the most blue- and red-shifted components at $V_{\rm LSR}\sim -119$~km~s$^{-1}$ and $\sim -17$~km~s$^{-1}$, respectively. The VLBA observations during 2003--2004 \citet{ima07b} found the most blue- and red-shifted components at $V_{\rm LSR}\sim -120$~km~s$^{-1}$ and $\sim-$10~km~s$^{-1}$, respectively. \cite{gom15} have already found a velocity spread ($V_{\rm LSR}\sim -125$ -- $\sim-$10~km~s$^{-1}$), that was similar to the one presented here. Thus the FLASHING spectra of the H$_2$O masers do not update the record of the jet's top speed. 

\subsection{Devolution of H$_2$O maser spectra towards the WFs}
\label{sec:result-devolution}

In some WFs, the H$_2$O maser spectrum has significantly changed, losing several components that restricted the total velocity spread to relatively narrow ranges. Some of the WFs showing this narrowing of velocity spread may be losing the high-velocity characteristics of H$_2$O maser spectrum of the WFs. 

In OH~12.8$-$0.9, the previous observations of the H$_2$O masers during 2004--2006 \citep{bob07} found the blue- and red-shifted components in the velocity ranges of $V_{\rm LSR}\sim-85$-- $-81$~km~s$^{-1}$ and $V_{\rm LSR}\sim-35$-- $-33$~km~s$^{-1}$, respectively. These authors also reported the evolution of the H$_2$O maser spectrum over 20 yr (1986--2007), in which the separation between the highest velocity components of the two clusters of spectral peaks had been extended by $\sim$12~km~s$^{-1}$. Through the five years of FLASHING, the blue-shifted cluster ($V_{\rm LSR}\sim-82$~km~s$^{-1}$) has been lost (figure \ref{fig:OH12.8-0.9}). Although a new component at $V_{\rm LSR}\sim -67$~km~s$^{-1}$ was tentatively detected, it was seen only once (on 2019 May 15) and the whole spectrum seems to have faded. 

In IRAS~18596$+$0315, the previous observations of the H$_2$O masers before 2004 June have found the blue- and red-shifted components around $V_{\rm LSR}\sim +$60 and $+$119~km~s$^{-1}$, respectively (e.g. \cite{dea07}). However, the double-horn profile of the spectrum has been often unstable since early observations (e.g., \cite{gom17}). Within five years of FLASHING, the blue-shifted components completely disappeared. 

\begin{figure}[ht]
\includegraphics[width=8.3cm]{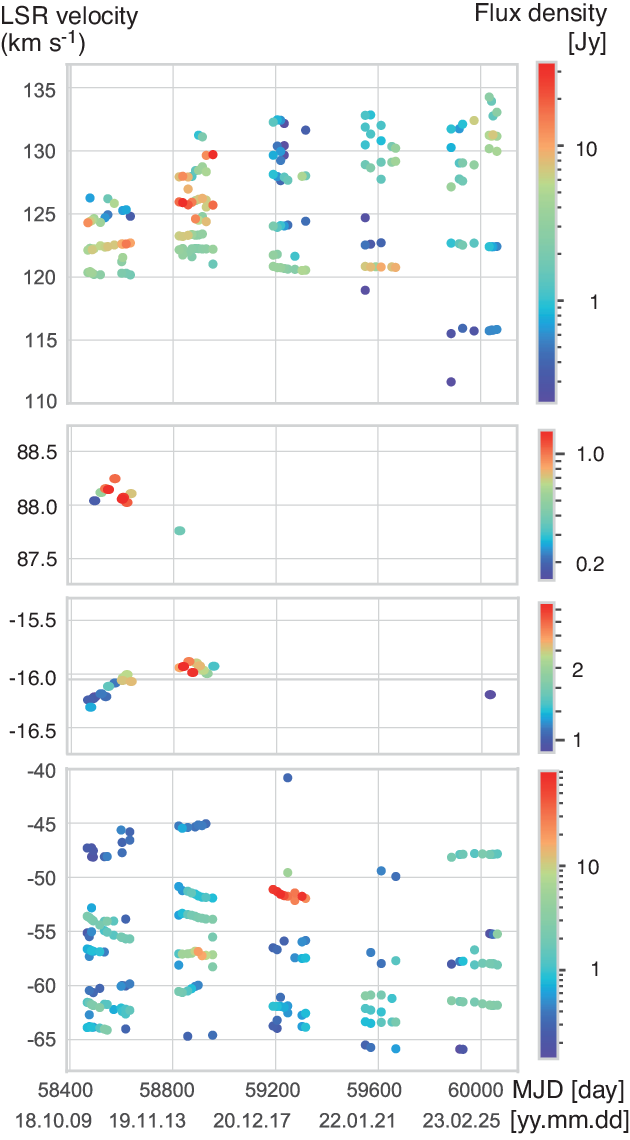} 
\caption{Dynamic spectra tracing spectral peaks of H$_2$O masers toward W~43A. The color of each data point indicates the flux density, following the scale shown in the right-side color bar.
The individual panels show groups of components in different LSR velocity ranges (shown in the left-hand axis).}
\label{fig:W43A-DS}
\end{figure}

\subsection{Systematic velocity drifts in W~43A}
\label{sec:result_vel-drifts} 

Velocity drifts of spectral peaks may be observed due to systematic acceleration and deceleration or local turbulence. One can see some examples of groups of line-of-sight velocity drifts of H$_2$O maser spectra towards long-period variables (e.g., \cite{lek99}). In the case of turbulence, each of the observed velocity drifts will be within a thermal velocity width ($\leq$1~km~s$^{-1}$) or that of microscopic turbulence often observed in VLBI observations (e.g., \cite{ima02b}) and they have a random trend. On the other hand, some systematic drifts should be found in a group of the spectral peaks if the masers are associated with accelerated or decelerated bulk motions, as seen in a Keplerian-rotating gas disk (e.g., \cite{has94}). 

Acceleration of maser regions is expected if they are associated with the gas in the CSE but entrained by the faster jet launched from the central stellar system. On the other hand, the deceleration is expected if they are in the jet decelerated by interaction with the ambient material in the CSE. It is difficult to determine which case can be considered for each velocity drift of the spectral peak, which is short-lived (one week--a few years). However, a preferential velocity drift is found in a velocity group of the spectral peaks, which may be associated with the individual gas clumps located closely to each other, it may indicate a specific dynamical trend of the gas clump group. This kind of spectral analysis is possible for the spectra that are rich in spectral peaks. 

Figure \ref{fig:W43A-DS} shows the dynamic spectra of H$_2$O masers tracing spectral peaks for W~43A. The individual peaks have a variety of drift directions. However, one can see preferential drift directions found commonly among a few peaks and repeatedly over the five years of the FLASHING observations. The red- and blue-shifted groups of the peaks have the preferential drift directions that indicate increasingly red- and blue-shifted velocities, respectively, both consistent with trends of acceleration with respect to the systemic stellar velocity ($V_{\rm LSR}\simeq$34~km~s$^{-1}$, \cite{taf20}).  

The dynamic spectra of H$_2$O masers tracking spectral peaks towards other WFs are shown in Appendix \ref{sec:dynamic-spectra}. For those sources, it is difficult to find any preferential direction of the grouped velocity drifts. It would be necessary to conduct more intensive monitoring observations (in a cadence of 1--2 weeks) for finding possible systematic drifts. Otherwise, too high condensation of the spectral peaks makes us difficult to precisely track the individual peaks from time to time. Even with the present data, for some sources, one can see that the velocity drifts, even if exist, are small ($<$1~km~s$^{-1}$ in five years). Note that such types of velocity variation we discuss here is still larger than the uncertainty of the velocities of the spectral peaks as described in Sect.\ \ref{sec:observations}. 

\begin{figure*}[ht]
\includegraphics[width=17cm]{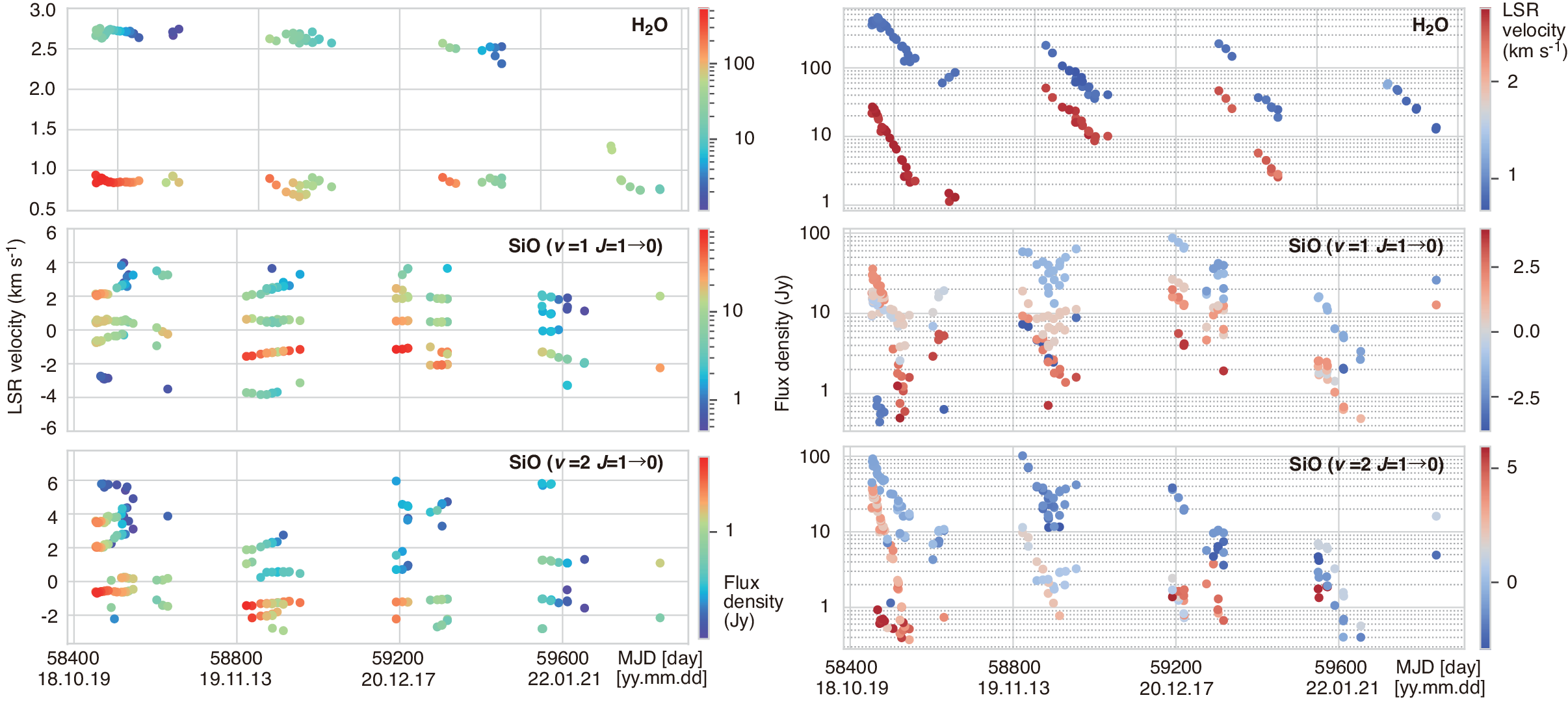} 
\caption{Time variation of the spectral peaks of the masers toward S~CrB. The top, middle, and bottom sub-panels show the variation of H$_2$O, SiO $v=1$, and SiO $v=2$ $J=1\rightarrow 0$ masers, respectively. 
The left and right sub-panels show the variation in LSR velocity and flux density for the individual spectral peaks, respectively.}
\label{fig:S-CrB-FD}
\end{figure*}

\begin{figure*}[ht]
\includegraphics[width=17cm]{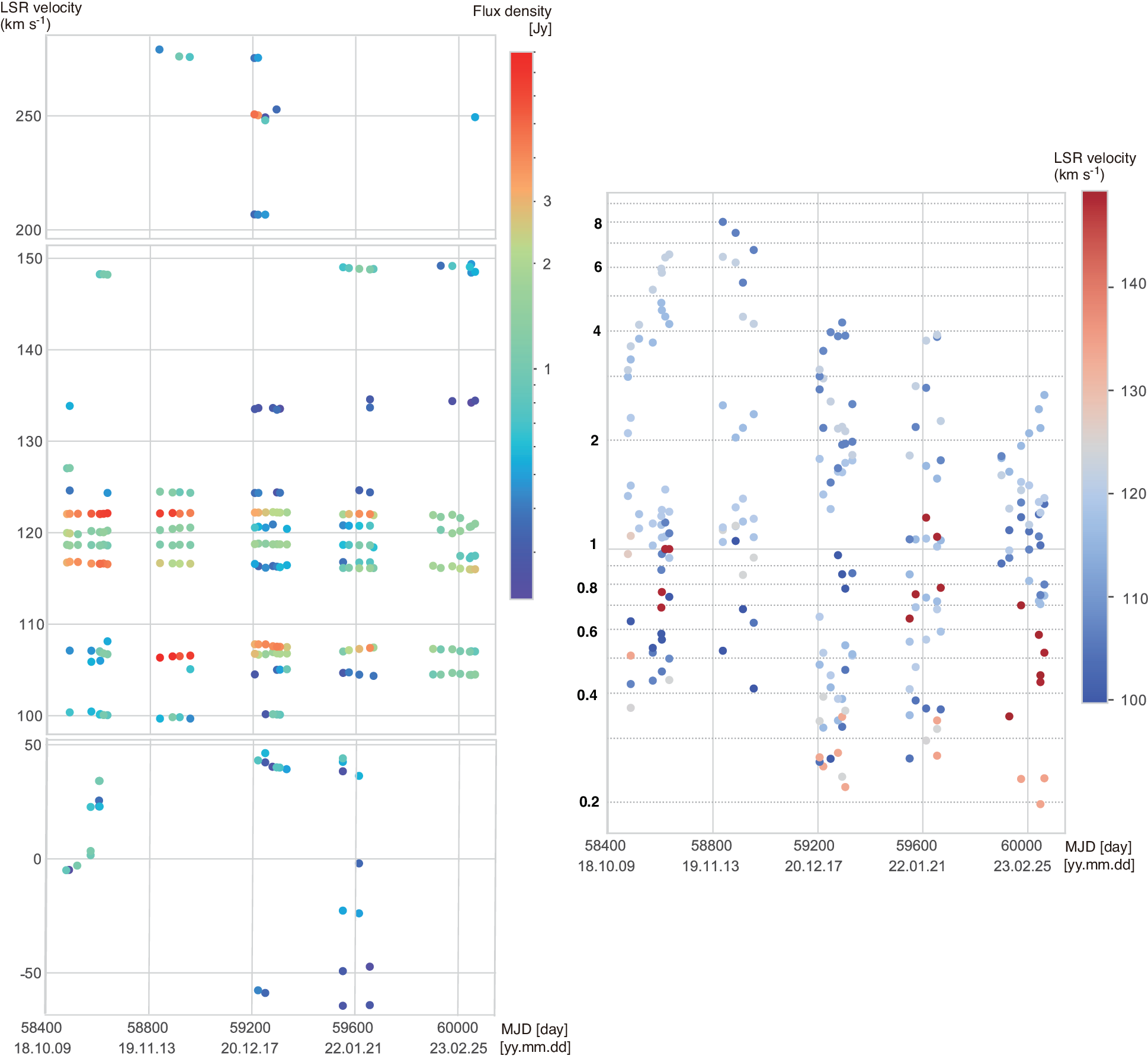}
\caption{Time variation of the individual spectral peaks of the IRAS~18460$-$0151 H$_2$O masers. Left and right panels show, respectively, the LSR domain and flux density domain variations of the spectral peaks.}
\label{fig:IRAS18460-0151-DS}
\end{figure*}

\subsection{Time variability of H$_2$O masers}
\label{sec:time-variability}

If some of the low-velocity components of H$_2$O masers are associated with a relic of the CSE as suggested by \citet{ima13b} and \citet{ima13c}, a periodic variation in their flux density is expected. In FLASHING, S CrB, a Mira variable with a stellar pulsation period of $\sim$360~d (General Catalog of Variable Stars, \cite{kho92}) had been observed in most of the FLASHING sessions until 2022 April, enabling us to test the technical feasibility for detecting such systematic variation. Figure \ref{fig:S-CrB-FD} shows the LSR velocities and flux densities of the spectral peaks of H$_2$O and SiO masers toward S~CrB. It demonstrates that FLASHING enables us to identify periodicity of the individual major and isolated spectral peaks in flux density variation when obtaining the data points every a few weeks during the observation seasons. 

The second and third panels from the top of figure \ref{fig:W43A-DS} show the time variation in LSR velocity and flux density for the individual spectral peaks in the low velocity components of the W~43A H$_2$O masers. In this source, two low-velocity components are found at $V_{\rm LSR}\sim-$16 and 88~km~s$^{-1}$, symmetrically in velocity with respect to the stellar systemic velocity ($V_{\rm LSR}\sim$34~km~s$^{-1}$, Sect.\ \ref{sec:result_vel-drifts}). They were observed roughly simultaneously, implying their common exciting source. However, it is difficult to identify any correlation in flux density variation between them.   

Figure \ref{fig:IRAS18460-0151-DS} shows the time variation the spectral peaks of the IRAS~18460$-$0151 H$_2$O masers, which host some low velocity components. Again, it is difficult to identify any correlation in flux density variation among the low-velocity components (99--148~km~s$^{-1}$). On the other hand, the blue-shifted ($V_{\rm LSR}<$50~km~s$^{-1}$) and the red-shifted ($V_{\rm LSR}>$200~km~s$^{-1}$) components roughly appeared simultaneously only during 2021 January--March. Thus it is difficult to explain the H$_2$O maser variation with the stellar origin, implying that the central star may be either at the end of the AGB phase or in the post-AGB phase when stellar pulsation is not found anymore. 

\subsection{Permanent death of SiO masers in W~43A}
\label{sec:result-W43A_SiO}

In FLASHING, no SiO maser detection has been reported in this WF. The stacking analysis of all observed spectra results in a negative detection at a 3-$\sigma$ level of $\sim$50~mJy. These masers had been constantly detected previously \citep{nak03,ima05}. The last detection of the masers was confirmed on 2009 April 18, but whose flux density had already been close to a level of tentative detection ($\sim$0.2~Jy, S. Deguchi, in private communication). In the data of the VLA observation published by \citet{yun17}, the SiO masers were not detected in 2013 June at a 3-$\sigma$ upper level of $\sim$0.02~Jy. Taking into account the stability of the SiO masers until the mapping observations with the VLA (in 2003 June, published by \cite{ima05}, and 2004 October and December), the negative detections over 14 years may indicate a permanent death, or at least significant decrease in the flux density of the maser by a factor of $\sim$10. 

\begin{figure}[ht]
\includegraphics[width=8.2cm]{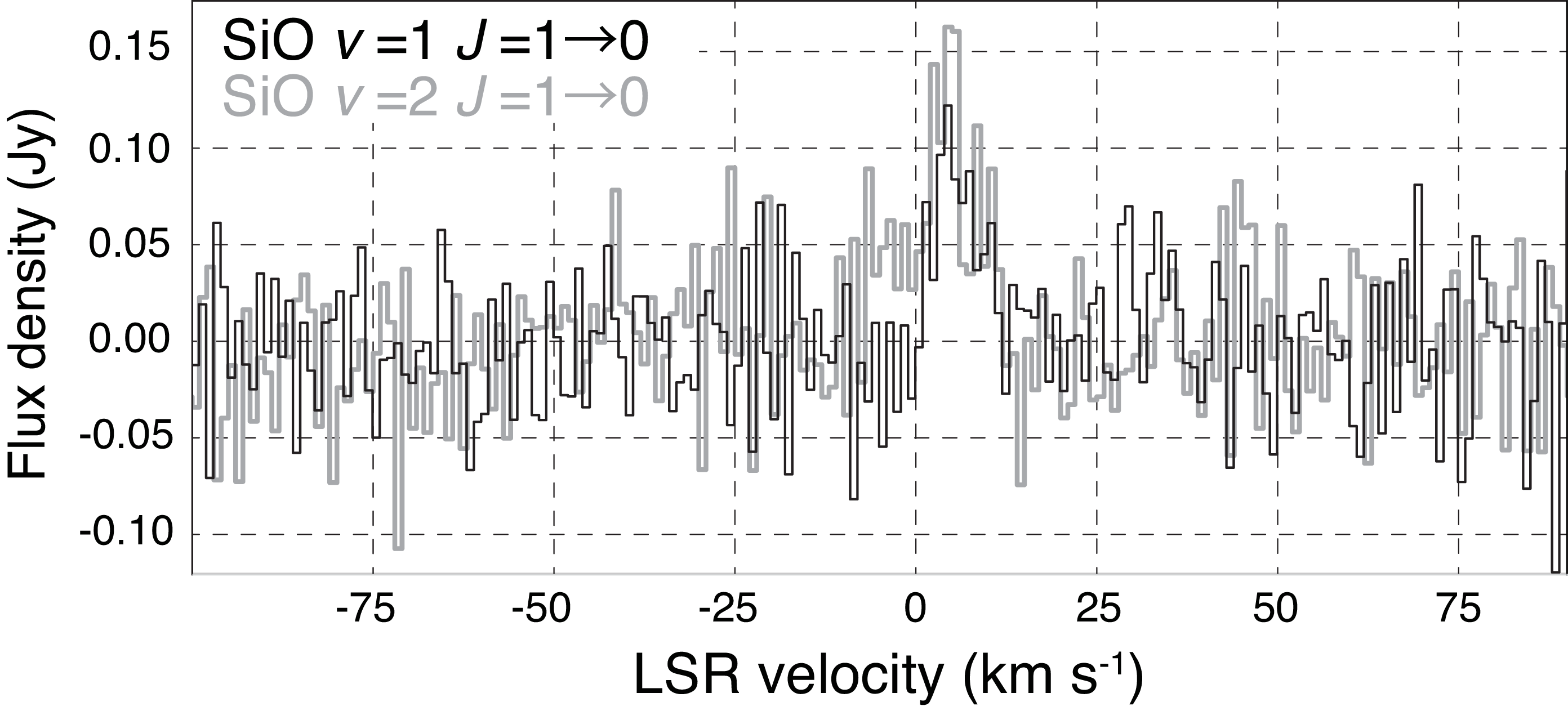} 
\caption{SiO $v=$1 (thick grey line) and 2 (thin black line) $J=1\rightarrow0$ maser spectra obtained by stacking all the data taken for IRAS16552$-$3050.}
\label{fig:IRAS16552-3050_SiOv1v2}
\end{figure}

\subsection{Evolution of SiO masers in IRAS~16552$-$3050}
\label{sec:result-IRAS16552-3050_SiO}

In the FLASHING project, SiO $J=1\rightarrow 0$ masers have been detected toward only IRAS~16552$-$3050 in the observed WFs. 

Paper II reported the first detections of SiO $J=1\rightarrow 0$ lines toward IRAS~16552$-$3050 in 2021 March, as the second case of SiO masers in the WFs after W~43A. In the data of this paper, these masers have been sometimes detected in the spectra with the velocity resolution of $\sim$0.4~km~s$^{-1}$, dependent on the noise level of the spectrum. Figures \ref{fig:IRAS16552-3050_SiOv1} and \ref{fig:IRAS16552-3050_SiOv2} 
in Appendix \ref{sec:spectra-SiO} show the time series of the SiO $v=$1 and 2 maser spectra toward IRAS~16552$-$3050. Paper II obtained clearer detections of the masers by smoothing the spectra shown here to a velocity resolution of $\sim$1.6~km~s$^{-1}$. \citet{ama23} also analyzed the FLASHING data at the latest epochs (2022 September--2023 April) in the same smoothed velocity resolution to confirm stable maser detections. 

In this paper, the spectra obtained by stacking all the spectral data are presented in figure \ref{fig:IRAS16552-3050_SiOv1v2}. Note that the dominant components of the masers were found around $V_{\rm LSR}\sim-5$~km~s$^{-1}$ in the year 2021 (Paper II) while those were around $V_{\rm LSR}\sim 5$~km~s$^{-1}$ through the whole FLASHING observations. Taking into account some variability of the maser emission, the spectral peaks in the stacked spectra imply their relatively stable appearance. Thus the SiO masers in IRAS~16552$-$3050 may exhibit basically a double-horn profile as seen in those in W~43A \citep{ima05}, suggesting that the SiO masers in IRAS~16552$-$3050 are also associated with be the base of the outflow exhibiting clear bipolarity revealed in H$_2$O masers \citep{sua08}.  

\begin{figure*}[ht]
\hspace*{-5mm}
\includegraphics[width=8.8cm]{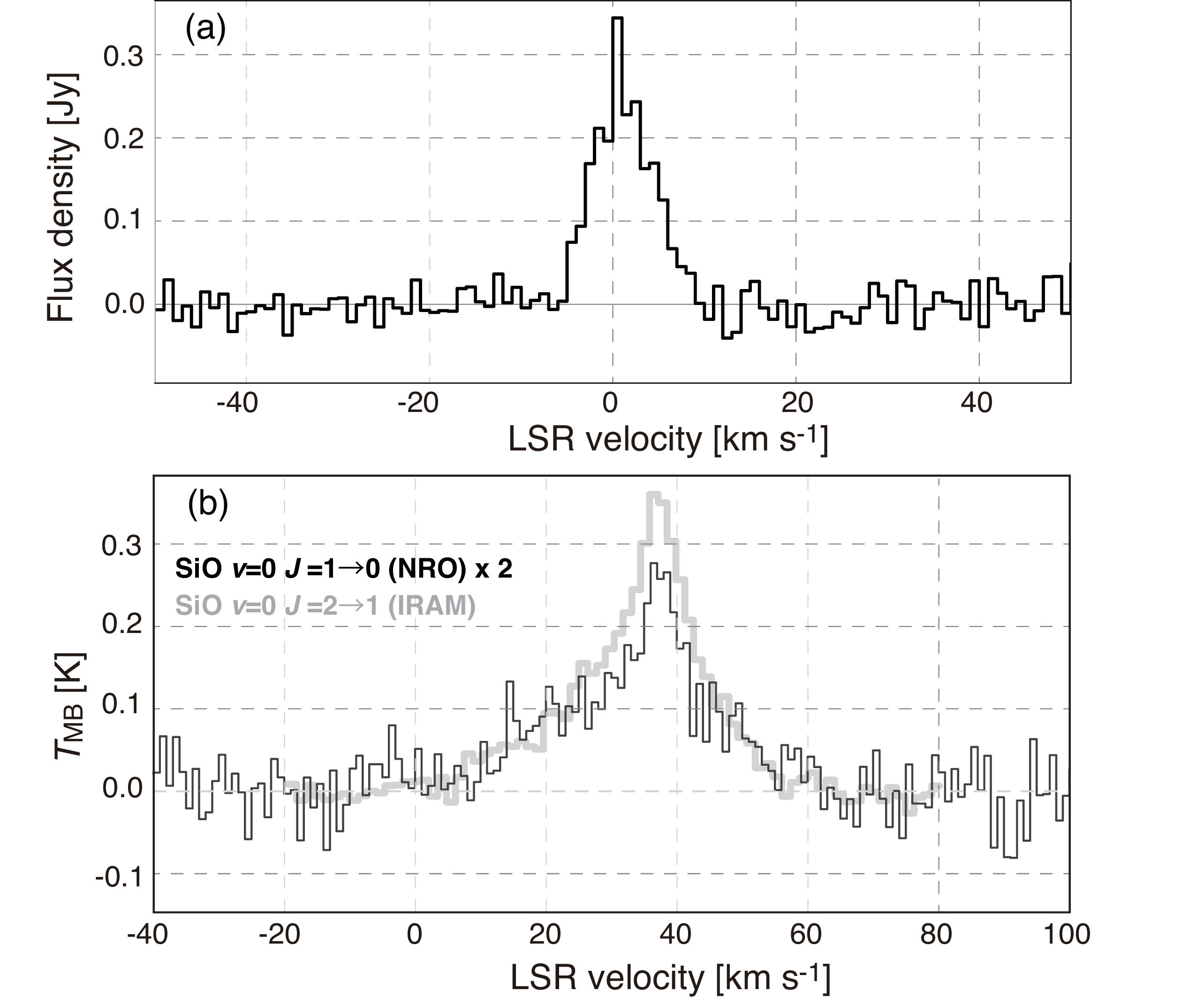} 
\caption{SiO $v=0$ $J=1\rightarrow0$ line spectra taken in FLASHING for (a) S CrB and (b) IRAS~19312$+$1950. For comparison, the spectrum of SiO $v=0\; J=2\rightarrow1$ line taken with the IRAM 30~m telescope \citep{qiu23} is superimposed in a thick gray line.}
\label{fig:SiOv0}
\end{figure*}

\subsection{Spectral stacking for the thermal SiO line}
\label{sec:result-SiOv0} 

For detecting faint emission of the ground vibrational-state ($v=0$) SiO $J=1\rightarrow 0$ line toward the individual target sources, all the FLASHING data were stacked for each source. This line may be composed of narrow maser and broad thermal components (e.g., \cite{dik21}). No such an emission line was detected over 3-$\sigma$ noise levels of 20--40~mJy towards the WFs. Instead, it was detected towards two out of the observed sources belonging to other categories. Figure \ref{fig:SiOv0} shows the spectra of the SiO $v=0$ line towards these sources. 

The first such case is S CrB (IRAS~15193$+$3132), a nearby ($\sim$400~pc) Mira variable, which we have observed for the system testing. We can see a slight asymmetry with some more emission at velocities red-shifted from the peak of the line profile. Note that the CORAVEL spectroscopic database gives the systemic line-of-sight velocity of $V_{\rm LSR}=-5.1\pm0.5$~km~s$^{-1}$ \citep{fam05}, while we derive it to be $V_{\rm LSR}=0.4\pm0.2$~km~s$^{-1}$ for the thermal component ($V_{\rm LSR}=0.0\pm0.2$~km~s$^{-1}$ for the maser component). 

The second one is IRAS~19312$+$1950, a peculiar source that is not clear whether it is an evolved star or a high-mass young stellar object. The deep integration of the FLASHING data enables us to detect the thermal SiO emission over a velocity range  comparable to that of H$_2$O masers ($\Delta v\sim$50~km~s$^{-1}$, \cite{deguchi04}). Interestingly, the line profile of the SiO $J=1\rightarrow 0$ quite resembles that of the $v=0$ $J=2\rightarrow 1$ line \citep{qiu23}. 

\section{Discussion}
\label{sec:discussion}

\subsection{Scientific prospects for monitoring observations of the WFs}
\label{sec:feasibility}

The scientific results of FLASHING, after five years of the monitoring observations, suggest that a higher cadence (roughly every week) may be useful for monitoring observations of the WFs, considering the properties of their variability, although realistic estimates of time availability at usable telescopes have to be taken into account.  

Although there exist only $\sim$15 WFs, it has been challenging to monitor all of them in the high cadence mentioned above. Most WFs are located at the coordinates close to the Galactic center, i.e., with a low elevation when observed from existing telescopes at the northern hemisphere, including Nobeyama. The targeted sensitivity to detect the extremely high velocity components and the ones associated with possible relics of AGB CSE should be as faint as $\sim$0.5~Jy so that they can be mapped in follow-up interferometric (especially VLBI) observations. Such sensitive radio telescopes are scarce in the southern hemisphere, at latitudes more favorable to observe these sources. 

The request for the higher cadence has been justified by the existence of short-lived (within a few weeks) extremely high velocity components that will break the top speed record of the jet in the excitation regions of H$_2$O masers. One can expect some periodicity of their appearance taking into account that expected in the ejections of the jet's material \citep{taf20}. The statistical data of the velocities and velocity drifts of such extreme components can be obtained if they are detected at several consecutive sessions. It is also expected that a single maser gas clump is found in H$_2$O maser spectra not only consecutively but also repeatedly. This possibility can be tested by such statistics. 

For some of the WFs exhibiting a forest of blended spectral peaks, such as IRAS~18286$-$0959, it is challenging to exactly track a specific spectral peak associated with the same maser gas clump from the observation session to another without continuously tracking change in its flux density. The case of a possible velocity drift in such a spectral peak also should be taken into account. 

This is a serious issue when identifying velocity components associated with a relic of a low-velocity CSE as expected (e.g., \cite{ima08,ima13b,ima13c}). If the central stellar system still hosts a long period variable star such as an AGB star, the maser emission should exhibit some periodicity caused by periodic change in the maser excitation condition in the CSE (e.g. \cite{shi08,kim19}). In the case of W~43A, this kind of behavior was confirmed in 1612~MHz OH masers \citep{her85}. It also have been found in H$_2$O and SiO masers, even in objects entering the post-AGB phase (e.g., \cite{kim19}). Because the observed sources have been monitored, in most cases, with a cadence at least monthly in the period from November or December to April or May, our time-series data analysis will be sensitive to periods longer than a few years. However, in the current status of our monitoring, identifying cases with such periodicity is still premature. 

The rate of secular growth of the H$_2$O maser spectrum of the WF, as seen in the history of record breaks of the jet's top speed found in the spectra (Sect.\ \ref{sec:result-evolution}) may indicate the dynamical timescale of the WF. For example, the emergence of new steady components as well as that of the short-lived, extremely high velocity components found in IRAS~18286$-$0959 has increased the expansion velocity of the jet in the steady components ($v_{\rm exp}$(steady)$\sim$240~km~s$^{-1}$) by $\sim$40~km~s$^{-1}$ in five years. This growth rate of the expansion velocity gives the dynamical timescale of the jet to be $\sim$30~yr, consistent with that previously estimated (Paper I). 

Thus a long-term monitoring is crucial to further test the hypothesis that the WFs are shortly lived but evolve within the lifetime of a human being ($\lesssim$100~yr). In a similar sense, firm identification of the WF's deathbed is also possible in the decadal monitoring. The WFs OH~12.8$-$0.9, IRAS~18596$+$0315 (Sect.\ \ref{sec:result-devolution}) and IRAS~19134$+$2131 (Sect.\ \ref{sec:result-evolution}) may be the ones which should be further monitored regularly in this aspect even at a lower cadence (every half to one year). The death of SiO masers in W~43A (Sect.\ \ref{sec:result-W43A_SiO}) also will be finally confirmed in the coming decade. 

The operation model of systematic monitoring observations of H$_2$O and SiO masers realized in FLASHING and its present outcome will possibly encourage a program that will be internationally coordinated so as to manage regular monitoring activities for the known WFs and the WF candidates. One of the examples is a successful synergy with ATCA observations (Paper III), in which the WFs in the southern sky have been independently and regularly observed, obtaining positional information of the spectral components of masers (in addition to their velocities and flux densities). This is useful to understand the evolutionary status of the WFs.

The capability of simultaneous SiO and H$_2$O maser observations is crucial for time efficient monitoring programs and higher chance of detections of SiO masers that will be newly ignited (Paper II) or revived although most of the WFs host only H$_2$O masers. This capability is useful especially for the WF candidates that now host both H$_2$O and SiO masers. 

\subsection{The life of the WFs} 
\label{sec:WF-live}

The theoretical model of H$_2$O maser excitation (e.g. \cite{hol13}) predicts two types of the maser excitation associated with C(continuous)- and J(jump)-type shocks, respectively, in which the gas density changes continuously and discontinuously from a pre- to post-shock gas layer. The C-type shocks are expected in the inner parts of typical CSEs with relatively high pre-shock densities ($n_{{\rm H}_{2}}
\sim 10^7$--$10^9$~cm$^{-3}$), while J-type ones in fast ($\geq$40~km~s$^{-1}$) outflows interacting with gas with a relatively low pre-shock density ($n_{{\rm H}_{2}}\sim 10^5$--$10^8$~cm$^{-3}$) as seen in star-forming regions. The two types of H$_2$O maser excitation are likely distinguishable in spectroscopic observations. More powerful masers are expected in the J-type shock excitation along the edge of the fast WF jet. Unstable properties of these masers may be explained by shock layers that are still being developed in a parental dense CSE whose gas density is too high for the J-type excitation but decreasing outward so that the J-type excitation becomes possible. However, because of blending of the spectral peaks, some of these J-type masers appear in anytime as low-velocity components even though they are unstable, together with more stable C-type masers. In such a situation, very high velocity components ejected by a bipolar jet may appear in the outer J-type shock layers and exhibit double peaks with clear symmetry with respect to the systemic velocity.  

On the other hand, note that the common envelope evolution (CEE) of a binary system is considered as a plausible model of the evolution of the WF system \citep{kho21}. Here it is interesting to explore when the CEE starts and ends. The parental CSE of the primary AGB/post-AGB star of the binary system will be destroyed by the jet generated by the companion star before or after these stars merge. For the case of a PN NGC~3132, which was imaged with the James Web Space Telescope \citep{dem22}, the first scenario is suggested. In this case, the CSE of the first evolved star would be destroyed by the jets of two companions out of four stars in the system before the companions were merged with the first star. Note that a binary system still exists after the previous merger of two stars if the stellar system was composed of these two and the third star. In this case, one can see a series of a fast evolution of the WF before merging with the third star. 

Without the third star that will approach and destroy the relic of the AGB CSE, one also can still expect to see such a relic even after a decease of the WF that formed before the first merger. The PN K3$-$35, which still hosts H$_2$O masers \citep{mir01}, may be in this case. From the total extent of the PN and a typical jet velocity in  PNs, the dynamical age of the PN K3$-$35 is estimated to be $\sim$800~yr. If the elongated PN is a remnant of the dead WF, one can consider a longer timescale of the WFs, comparable to this PN's age. 

Taking into account the models of H$_2$O maser excitation and the CEE in the WFs, here we speculate the scenario of a life of the WF source as follows. Note that \citet{vin04} also considered the scenario of the early phase of the WF evolution, but without the discussion taking into account the existence of a binary system. 

The launch of a WF jet is expected at the end of the AGB phase, such as the phase of an OH/IR star. In this stage, the CSE is significantly developed so that 1612~MHz OH masers are generated while H$_2$O masers are being quenched due to a too high gas density. Because of the large size of the CSE, the companion is swallowed in the CSE and generates the initial jet. If the jet launch starts earlier, one can see more AGB stars that host H$_2$O masers whose spectrum should be rich in spectral peaks within a typical expansion velocity of the AGB CSEs ($v_{\rm exp}=$10--30~km~s$^{-1}$). Actually, except a few cases including IRAS~18286$-$0959 and IRAS~18460$-$0151, the H$_2$O maser spectra of the WFs are not so rich in such low-velocity components.  

The growth of a WF jet is expected until the expansion velocity found in the H$_2$O maser spectrum reaches that traced in thermal CO emission as seen in W~43A. H$_2$O masers in W~43A were excited in the middle way of the bipolar jet in the year 2017, where the CSE material is entrained by the jet \citep{taf20}, while they will achieve the tips of the jet so as to have the highest expansion velocity mentioned above. Another WF in this stage may be IRAS~18113$-$2503; both this source and W~43A exhibit clear periodic patterns in the distributions of H$_2$O maser features along the jets \citep{cho15,oro19,taf20}. 

It will take $\lesssim$50~yr, half of the total lifetime of the WF, to go from the first to the second stages mentioned above, the WFs in the second stages would have experienced several cycles of the binary orbital motions as indicated by the maser distribution patterns mentioned above. Between the two stages, the property of a collimated bipolar jet would be developed. IRAS~16552$-$3050, IRAS~18286$-$0959, and IRAS~19190$+$1102 may be in this phase, in which the top speed of the jet indicated by H$_2$O masers increases. In the former, SiO masers would be ignited with increase in the mass loss rate by the jet (Paper II). 

At the late stage of the WF evolution, the double-horn profiles of the H$_2$O maser spectra are decaying because of the intrinsic asymmetry in the location of the companion star, or the driving source of the jet with respect to the CSE and complete penetration of the jet out to the CSE. The WFs in the devolution phase (Sect.\ \ref{sec:result-devolution}) may correspond to this phase. Some of them may get visible in optical, such as IRAS~19134$+$2131 \citep{ima07b}, and exhibit photo-ionization, such as IRAS~18043$-$2116 whose jet's speed traced by H$_2$O masers also achieves that by CO emission (Paper III).

Based on the comparison of the morpho-spectral data of H$_2$O masers towards the WFs in FLASHING and the interferometric observations, we have discussed the evolution of the WFs as describe above. Interestingly, we will be able to see and soon test the transition from one to another phase of the WF evolution within our lifetime.  This should engage in the monitoring observation projects described in Sect.\ \ref{sec:feasibility}. 

Note also that other H$_2$O-maser-emitting post-AGB stars, when first discovered, may not look like the WFs, but they might be. Therefore, only with surveys it is not possible to estimate the present-day total number of the WFs. Anyway, exactly estimating the number of the WFs is crucial through monitoring observations for the existing WFs and new surveys for WFs, after an interesting statistics by \citet{kho21} to find out whether all ``common envelope" sources would turn out to be the WFs at some point of their evolution path. One of the ideas for the new WF surveys is to target circumstellar OH maser sources detected in deep unbiased sky surveys such as the Galactic ASKAP Spectral Line Survey (GASKAP, \cite{dic13}). 

\begin{ack}
We acknowledge Yuri Uno, Yosuke Shibata, Miki Takashima, Daichi Maeyama, and Ryosuke Yamaguchi for their  contributions to the observation operation and the data reduciton. 

The Nobeyama 45-m radio telescope is operated by Nobeyama Radio Observatory, a  branch of National Astronomical Observatory of Japan (NAOJ), National Institutes of Natural Sciences. HI and GO are supported by JSPS KAKENHI Grant Number JP16H02167. JFG acknowledges support from grant Nos. PID2020-114461GB-I00 and CEX2021-001131-S, funded by MCIN/AEI/10.13039/50110001103. HI and JFG were supported by the Invitation Program for Foreign Researchers of the Japan Society for Promotion of Science (JSPS grant S14128). GO was supported by the Australian Research Council Discovery project DP180101061 of the Australian government, and the grants of CAS LCWR 2018-XBQNXZ-B-021 and National Key R\&D Program of China 2018YFA0404602. DT was supported by the ERC consolidator grant 614264. RB acknowledges support through the EACOA Fellowship from the East Asian Core Observatories Association. LU acknowledges support from the University of Guanajuato grant No. CIIC 090/2023. 
\end{ack}

\appendix

\section{Logs of the FLASHING observations}
\label{sec:log}

Table \ref{tab:obslogs} gives the log of the FLASHING observations for the sources listed in table \ref{tab:sources} to show the dates of the observations and the records of positive and negative detections of the masers. The date format is YY-MM-DD (year-month-day).

The data of the maser source spectra are available in ASCII text format by contacting the first author of this paper.

\renewcommand{\arraystretch}{0.9}
\begin{longtable}{l
c@{\hspace*{-3pt}}
c@{\hspace*{-3pt}}
c@{\hspace*{-3pt}}
c@{\hspace*{-3pt}}
c@{\hspace*{-3pt}}
c@{\hspace*{-3pt}}
c@{\hspace*{-3pt}}
c@{\hspace*{-3pt}}
c@{\hspace*{-3pt}}
c@{\hspace*{-3pt}}
c@{\hspace*{-3pt}}
c@{\hspace*{-3pt}}
c@{\hspace*{-3pt}}
c@{\hspace*{-3pt}}
c@{\hspace*{-3pt}}
c@{\hspace*{-3pt}}
c@{\hspace*{-3pt}}
c@{\hspace*{-3pt}}
c@{\hspace*{-3pt}}
c@{\hspace*{-3pt}}
c@{\hspace*{-3pt}}
c@{\hspace*{-3pt}}
c@{\hspace*{-3pt}}
}
    \caption{Observation log for the maser sources monitored in FLASHING}
    \label{tab:obslogs} \\ \hline
     Date &        a &        b &        c &        d &        e &        f &        g &        h &        i &        j & k &        l &        m &        n &        o &        p &        q &        r &        s &        t &        u &        v \\ \hline
    \endfirsthead
    \hline
    Date &        a &        b &        c &        d &        e &        f &        g &        h &        i &        j &  k &        l &        m &        n &        o &        p &        q &        r &        s &        t &        u &        v \\ \hline
    \endhead
    \hline
    \endfoot
    \hline              
\multicolumn{23}{l}{--:  No observation; $\times$: Negative maser detection;
B: Detections of both H$_2$O and SiO masers; H: Detection of H$_2$O maser.} \\
   \endlastfoot
        18-12-01 & B & -- & -- & -- & -- & -- & -- & -- & -- & -- & -- & -- & -- & -- & -- & -- & -- & -- & -- & -- & -- & -- \\
                18-12-04 & -- & -- & -- & -- & -- & -- & -- & -- & -- & -- & -- & -- & -- & -- & -- & -- & -- & H & -- &  -- & -- & -- \\
        18-12-06 & -- & -- & -- & -- & -- & -- & -- & -- & -- & -- & -- & -- & -- & -- & -- & -- & -- &        H & -- & -- & -- & -- \\
                18-12-07 & -- & -- & -- & -- & -- & -- & -- & -- & -- & -- & -- & -- & H & -- & -- & -- & -- & -- & H & -- & -- & -- \\
        18-12-08 & B & -- & -- & -- & -- & -- & -- & -- & -- & -- & -- & -- & -- & -- & H & -- & -- & -- & -- & -- & -- & -- \\        
                18-12-09 & B & -- & -- & -- & -- & -- & -- & -- & -- & B & -- & -- & -- & -- & -- & -- & -- & -- & -- & -- & -- & -- \\           18-12-10 &  --  & -- & -- & -- & -- & -- & -- & -- & -- &  --  & -- & -- & -- & 
                -- & -- & -- & -- & H & H & -- & -- & $\times$ \\                
        18-12-12 & -- & -- & -- & -- & -- & -- & -- & -- & -- & -- & -- & -- & -- & -- & -- &        H & -- & -- & -- & -- & -- & -- \\
        18-12-13 & -- & -- & -- & -- & -- & -- & -- & -- & -- & -- & -- & -- & -- & -- & -- & -- & -- & H & -- & -- & -- & -- \\
        18-12-15 & -- & -- & -- & -- & -- & -- & -- & -- & -- & -- & -- & -- &        H & -- & -- & -- & -- & -- & -- & -- & -- & -- \\
        18-12-20 & B & -- & -- & -- & -- & -- & -- & -- & -- & -- & -- & -- & -- & -- & -- & -- & -- & -- &        H & -- & -- & -- \\
        18-12-22 & -- & -- & -- & -- & -- & -- & -- & -- & -- & -- & -- & -- & -- & -- & -- & -- & -- &        H & -- & -- & -- & -- \\
        18-12-23 &        B & -- & -- & -- & -- & -- & -- & -- & -- & -- & -- & -- &        H &        H &        H & -- & -- & -- & -- & -- & -- & -- \\
        18-12-29 & -- & -- & -- & -- & -- & -- & -- & -- & -- & -- & -- & -- & -- & -- & -- & -- & -- & -- & -- & -- & -- &        $\times$ \\
        18-12-30 &        B & -- & -- & -- & -- & -- & -- & -- & -- & -- & -- & -- & -- & -- & -- & -- & -- & -- & -- & -- & -- &        $\times$ \\
        18-12-31 & -- & -- & -- & -- & -- & -- & -- & -- & -- & -- & -- & -- & -- & -- & -- &        H & -- & -- & -- & -- & -- & -- \\
        19-01-01 & -- & -- & -- & -- & -- & -- & -- & -- & -- & -- & -- & -- &        H & -- & -- & -- & -- & -- & -- & -- & -- & -- \\
        19-01-02 & -- & -- & -- & -- & -- & -- & -- &        H & -- &        B & -- & -- & -- & -- & -- & -- & -- & -- & -- & -- & -- & -- \\
        19-01-03 & -- &        H & -- & -- & -- & -- & -- & -- & -- & -- & -- & -- & -- & -- &        H & -- & -- & -- &        H & -- & -- & -- \\
        19-01-04 & -- & -- & -- & -- & -- & -- & -- & -- &        $\times$ & -- & -- &        B & -- &        H & -- & -- & -- & -- & -- & -- & -- & -- \\
        19-01-05 & -- & -- & -- & -- & -- & -- & -- & -- & -- & -- &        H & -- & -- & -- & -- & -- & -- &        H & -- & -- & -- & -- \\
        19-01-08 & -- & -- & -- & -- & -- & -- & -- & -- & -- & -- & -- & -- &        H & -- & -- &        H & -- & -- & -- & -- & -- & -- \\
        19-01-22 & -- & -- & -- & -- & -- & -- & -- & -- & -- & -- & -- & -- & -- & -- & -- & -- & -- & -- &        H & -- & -- & -- \\
        19-02-02 &        B & -- & -- & -- & -- & -- & -- & -- & -- & -- & -- & -- &        H & -- & -- & -- & -- & -- & -- & -- & -- & -- \\
        19-02-04 & -- & -- & -- & -- & -- & -- & -- & -- & -- & -- & -- & -- & -- &        $\times$ &        H & -- & -- &        H & -- & -- & -- & -- \\
        19-02-06 & -- & -- & -- & -- & -- & -- & -- & -- & -- & -- &        H & -- & -- & -- & -- & -- & -- & -- & -- & -- & -- &        $\times$ \\
        19-02-09 & -- & -- & -- & -- & -- & -- &        H & -- & -- &        B & -- & -- & -- & -- & -- & -- & -- & -- & -- & -- & -- & -- \\
        19-02-12 & -- & -- & -- & -- & -- & -- & -- & -- & -- & -- & -- & -- & -- & -- & -- &        H & -- & -- & -- & -- & -- & -- \\
        19-02-16 &        B & -- & -- & -- & -- & -- & -- & -- & -- & -- & -- &        B & -- & -- & -- & -- & -- & -- &        H & -- & -- & -- \\
        19-02-20 &        B &        H & -- & -- & -- & -- & -- & -- & -- & -- & -- & -- &        H & -- & -- & -- & -- & -- & -- & -- & -- & -- \\
        19-03-03 &        B & -- & -- & -- & -- & -- & -- & -- & -- & -- & -- & -- &        H & -- & -- & -- & -- & -- & -- & -- & -- & -- \\
        19-03-06 & -- &        H & -- & -- &        H & -- & -- & -- & -- & -- &        H & -- & -- & -- & -- & -- & -- &        H & -- & -- & -- & -- \\
        19-03-17 & -- & -- & -- & -- & -- & -- & -- & -- & -- & -- & -- &        B & -- & -- & -- & -- & -- & -- & -- & -- & -- &        $\times$ \\
        19-03-21 & -- & -- & -- &        H & -- &        H & -- & -- & -- & -- & -- &        H & -- & -- & -- & -- & -- & -- &        H & -- & -- &        $\times$ \\
        19-03-28 & -- & -- & -- & -- & -- & -- & -- & -- & -- & -- & -- & -- &        H & -- &        H & -- & -- & -- & -- & -- & -- & -- \\
        19-04-15 & -- & -- & -- & -- & -- & -- & -- & -- & -- & -- &        H & -- & -- & -- & -- & -- & -- & -- & -- & -- & -- & -- \\
        19-04-26 & -- & -- & -- & -- & -- & -- & -- & -- & -- & -- & -- & -- &        H & -- & -- &        $\times$ & -- &        H & -- & -- & -- & -- \\
        19-04-27 & -- &        H &        $\times$ & -- & -- & -- & -- & -- &        $\times$ &        B & -- &        B & -- &        $\times$ & -- & -- & -- & -- & -- & -- & -- & -- \\
        19-04-28 & -- & -- & -- & -- & -- & -- & -- & -- & -- & -- &        H & -- & -- & -- & -- & -- & -- & -- &        H & -- & -- &        $\times$ \\
        19-04-29 &        B & -- & -- & -- & -- & -- &        H &        H & -- & -- & -- & -- &        H & -- &        H & -- & -- & -- & -- & -- & -- & -- \\
        19-04-30 & -- & -- & -- & -- &        H &        H & -- &        H & -- & -- & -- & -- &        H & -- & -- & -- & -- & -- & -- & -- & -- & -- \\
        19-05-02 & -- & -- & -- &        H &        H &        H & -- & -- & -- & -- & -- & -- & -- & -- &        H & -- & -- & -- & -- & -- & -- & -- \\
        19-05-14 &        B & -- & -- & -- & -- & -- & -- & -- & -- &        B &        H &        B &        H & -- &        H & -- & -- & -- & -- & -- & -- & -- \\
        19-05-15 & -- & -- & -- & -- & -- & -- & -- & -- &        H & -- & -- & -- & -- & -- & -- &        H & -- & -- & -- & -- & -- & -- \\
        19-05-24 & -- & -- & -- & -- & -- & -- & -- & -- & -- & -- & -- & -- & -- & -- & -- & -- & -- & -- & -- & -- & -- &        $\times$ \\
        19-05-25 & -- & -- &        $\times$ & -- & -- & -- & -- & -- & -- & -- & -- & -- & -- &        $\times$ & -- & -- & -- &        H & -- & -- & -- & -- \\
        19-05-26 & -- & -- & -- & -- & -- & -- & -- & -- & -- & -- & -- & -- & -- & -- & -- & -- & -- &        H & -- & -- & -- & -- \\
        19-05-27 &        B & -- & -- & -- & -- & -- & -- & -- & -- &        B & -- & -- & -- & -- & -- & -- & -- & -- & -- & -- & -- & -- \\
        19-05-28 & -- & -- & -- &        H & -- & -- & -- & -- & -- & -- & -- & -- & -- & -- & -- & -- & -- & -- &        H & -- & -- & -- \\
        19-05-29 & -- & -- & -- & -- &        H & -- & -- & -- & -- & -- &        H & -- & -- & -- & -- & -- & -- & -- &        H & -- & -- & -- \\
        19-05-30 & -- &        H & -- & -- & -- & -- &        H & -- & -- & -- & -- & -- &        H & -- &        H &        H & -- & -- & -- & -- & -- & -- \\
        19-05-31 & -- & -- & -- & -- & -- &        H & -- &        H & -- & -- & -- & -- & -- & -- & -- & -- & -- & -- & -- & -- & -- & -- \\
                19-12-06 & B & H & -- & -- & -- & -- & -- & -- & -- & -- & H & -- & H & -- & -- & -- & -- & 
                H & -- & -- & -- & -- \\                
        19-12-20 &        B & -- & -- & -- & -- & -- & H & -- & -- & -- &  H & -- & H &  -- &      H & -- & -- &
        -- & -- & H & -- & --  \\
        20-01-10 &        B & -- & -- & -- & -- & -- & -- & -- & -- & -- &        H & -- &        H & -- & -- & -- & -- & -- & -- & -- & -- & -- \\
        20-01-25 &        B & -- & -- & -- & -- & -- & -- & -- & -- & -- &        H & -- &        H & -- & -- & -- & -- &        H & -- &        H & -- & -- \\
        20-02-07 &        B &        H & -- & -- & -- & -- & -- & -- & -- & -- &        H & -- & -- & -- &        H & -- & -- & -- &        H & -- & -- & -- \\
        20-02-08 &        B & -- & -- & -- & -- & -- &        H & -- & -- &        B & -- &        B &        H & -- & -- & -- & -- & -- & -- &        H & -- & -- \\
        20-02-20 &        B & -- & -- & -- & -- & -- & -- & -- &        H & -- &        H &        B &        H & -- & -- & -- & -- & -- & -- &        H & -- & -- \\
        20-02-22 &        B & -- &        $\times$ & -- &        $\times$ & -- & -- & -- & -- &        H & -- & -- & -- & -- &        H & -- & -- &        H &        H & -- & -- & -- \\
        20-03-06 &        B & -- & -- & -- & -- &        H & -- & -- &        H &        B &        H & -- &        H & -- &        H & -- & -- & -- & -- &        H & -- & -- \\
        20-03-07 &        B &        H & -- & -- & -- & -- &        H &        H & -- & -- & -- &        B & -- & -- & -- &        H & -- &        H &        H & -- & -- & -- \\
        20-03-19 &        B & -- & -- &        H & -- &        H & -- & -- & -- &        B &        H & -- & -- & -- & -- & -- & -- & -- &        H &        H & -- & -- \\
        20-03-21 &        B &        H & -- & -- & -- & -- & -- & -- & -- & -- & -- &        B &        H &        $\times$ & -- &        H & -- &        H & -- & -- & -- & -- \\
        20-04-16 &        B &        H &        $\times$ & -- &        $\times$ &        H &        H & -- &        H &        B &        H &        B &        H & -- &        H &        H & -- &        H &        H &        H & -- &        $\times$ \\

        20-12-07 & -- & -- & -- & -- &  -- & H & H & -- & H & H & -- & B & H & -- & -- &  -- &  -- & -- & H &  -- & -- & $\times$ \\

        20-12-08 &        B & -- & -- & -- &        $\times$ & -- & -- &        H & -- & -- &        H & -- & -- & -- &        H &        H &        S &        H & -- &        H &        B & -- \\
        20-12-23 &        B &        H & -- & -- & -- & -- &        H & -- & -- &        B &        H &        B &        H &        $\times$ &        H & -- & -- & -- & -- & -- &        H & -- \\
        21-01-05 &        B & -- & -- &        $\times$ & -- &        H &        H & -- & -- &        B & -- &        B &        H &        $\times$ & -- & -- & -- & -- &        H &        H & -- &        $\times$ \\
        21-01-06 &        B &        H & -- & -- & -- & -- & -- & -- & -- & -- &        H & -- & -- & -- &        H &        H &        S &        H & -- &        H &        B & -- \\
        21-01-19 &        $\times$ &        H & -- & -- & -- &        H &        H & -- &        H &        B &        H &        B &        H & -- & -- & -- & -- & -- & -- & -- & -- & -- \\
        21-02-02 & -- & -- & -- & -- & -- & -- & -- & -- & -- &        B & -- &        B &        H & -- & -- & -- & -- & -- &        H & -- & -- & -- \\
        21-02-03 & -- & -- & -- &        $\times$ & -- &        H & -- &        H & -- & -- &        H & -- & -- & -- &        H & -- &        S &        H & -- & -- & -- & -- \\
        21-03-02 &        B & -- & -- & -- & -- &        H & -- & -- &        H &        B &        H &        B &        H & -- & -- & -- & -- &        H &        H & -- & -- & -- \\
        21-03-03 &  -- & -- & -- & -- & -- & -- &        H &        -- & -- & -- & -- & -- & -- & -- &        H &        H &        S & -- & -- &        H &        B & -- \\
        21-03-19 &        B &        H & -- & -- &        H &        H &        H &        H & -- &        B &        H &        B &        H & -- &        H & -- & -- & -- &        H & -- & -- &        $\times$ \\
        21-03-30 &        B & -- & -- & -- & -- & -- &        H & -- & -- &        B & -- &        B &        H & -- & -- &        H &        S & -- & -- &        H & -- &        $\times$ \\
        21-03-31 &        B &        B & -- &        $\times$ & -- &        H & -- &        H & -- & -- &        H & -- & -- & -- &        H & -- & -- &        H &        H & -- &        B & -- \\
        21-04-13 &        B &        B & -- & -- & -- & -- &        H & -- &        H &        B &        H &        B &        H & -- & -- & -- & -- & -- &        H &        H & -- & -- \\
        21-04-26 &        B & -- & -- & -- & -- & -- & -- & -- & -- & -- & -- & -- &        H & -- & -- & -- & -- & -- & -- & -- & -- & -- \\
        21-04-27 & -- &        B & -- & -- & -- &        H & -- &       -- & -- &        B &        H &        B & -- & -- &        H &        H & -- &        H &        H & -- &        B & -- \\
       
              21-12-01 &  B & H & -- & -- & -- & -- & -- & -- &  -- & B & -- & B & H & -- & H & -- & -- & H & -- & H & B & -- \\
        
        21-12-03 &        B &        $\times$ & -- & -- & -- &        H & -- & -- &        H &        B &        H &        B &        H & -- & -- & -- & -- & -- &        H & -- & -- &        H \\
        21-12-22 &        B &        H & -- & -- & -- & -- & -- & -- & -- &        B &        H & -- &        H & -- & -- &        $\times$ & -- &        H & -- &        H & -- & -- \\
        21-12-24 &        B &        $\times$ & -- & -- & -- & -- & -- & -- & -- &        H & -- & -- &        $\times$ &        $\times$ &        H & -- &        S & -- &        H & -- &        S & -- \\
        22-01-11 &        B & -- & -- & -- & -- &        H & -- & -- & -- & -- & -- & -- &        H & -- & -- & -- & -- &        $\times$ & -- & -- & -- & -- \\
        22-02-01 &        B &        H & -- & -- & -- &        H & -- &        H &        H &        B & -- &        B &        H & -- & -- & -- & -- &        H & -- & -- &        B & -- \\
        22-02-02 &        B &        S & -- &        $\times$ &        $\times$ & -- &        H & -- & -- & -- &        H & -- &        $\times$ & -- &        H &        $\times$ & -- & -- &        H &        H & -- & -- \\
        22-03-15 &        B & -- & -- & -- & -- &        H & -- & -- &        $\times$ &        B &        H &        B &        H & -- & -- &        $\times$ & -- &        H & -- &        H & -- & -- \\
        22-03-16 &        B &        B & -- & -- & -- & -- &        H &        H &        $\times$ & -- & -- & -- & -- & -- &        H & -- &        S & -- & -- &        H &        B & -- \\
        22-03-30 & -- &        H & -- & -- & -- &        H &        H &        H & -- &        B &        H &        B &        H & -- &        H & -- & -- &        H &        H &        H &        S & -- \\
        22-09-17 &        B &        H & -- & -- & -- &        H & -- & -- & -- & -- & -- & -- & -- &        $\times$ & -- & -- & -- &        H &        H & -- & -- & -- \\
        22-09-18 & -- & -- & -- & -- & -- & -- &        H & -- &        $\times$ & -- &        H & -- &        H & -- &        H & -- & -- & -- & -- & -- &        S &        $\times$ \\

         22-11-01 & B &  -- & -- & -- & $\times$ & H & -- & H & -- & 
         B & H & S & H & -- & -- & -- & S & -- & -- & -- & -- & -- \\
        
        22-11-16 & -- &        H & -- &        H & -- & -- &        H & -- & -- &        B &        H & -- &        H & -- &        H & -- & -- &        H & -- & -- & -- & -- \\
        22-12-03 & -- & -- & -- & -- & -- &        H & -- &        H &        $\times$ &        B &        H &        S &        H & -- & -- & -- & -- & -- &        H & -- & -- & -- \\
        22-12-16 & -- & -- & -- & -- & -- & -- & -- & -- & -- &        B &        H &        S &        H & -- &        H &        $\times$ &        S & -- & -- & -- & -- &        $\times$ \\
        23-01-05 & -- &        B & -- &        $\times$ &        $\times$ & -- & -- & -- & -- &        B & -- &        $\times$ &        H &        $\times$ &        H & -- & -- &        $\times$ & -- &        $\times$ & -- & -- \\
        23-01-17 & -- &        H & -- & -- & -- &        H &        H &        H & -- &        B &        H & -- & -- & -- & -- & -- &        S & -- &        H & -- &        B & -- \\
        23-01-30 & -- & -- & -- & -- & -- & -- & -- & -- &        H &        B &        H &        S &        H & -- &        H &        $\times$ & -- &        $\times$ & -- & -- & -- & -- \\
        23-03-02 & -- &        H & -- &        $\times$ &        $\times$ &        H &        H &        H &        H &        B &        H &        S &        H &        $\times$ &        H & -- &        S & -- &        H &        $\times$ & -- & -- \\
        23-03-30 & -- & -- & -- &        $\times$ & -- &        H & -- &        H &        H & -- &        H &        S &        H & -- & -- & -- & -- & -- &        H & -- & -- & -- \\
        23-04-08 & -- &        B & -- & -- & -- & -- &        H & -- & -- &        B & -- & -- &        H & -- &        H &        H &        B &        $\times$ & -- &        $\times$ & -- &        $\times$ \\
        23-04-15 & -- &        H & -- & -- & -- & -- &        H &        H & -- &        B &        H &        S &        H & -- &        H & -- & -- & -- & -- & -- & -- & -- \\
        23-04-22 & -- & -- & -- & -- & -- & -- & -- & -- &        H &        B & -- & -- & -- &        $\times$ & -- & -- &        B & -- &        H & -- &        S & -- \\
        23-04-29 & -- &        B & -- & -- & -- &        H & -- &        H & -- &        B &        H &        S &        H & -- &        H &        $\times$ &        S & -- & -- & -- & -- & -- \\
\end{longtable}

\section{Time series of H$_2$O maser spectra taken for the WFs}
\label{sec:spectra}

\begin{figure*}[ht]
\includegraphics[width=17cm]{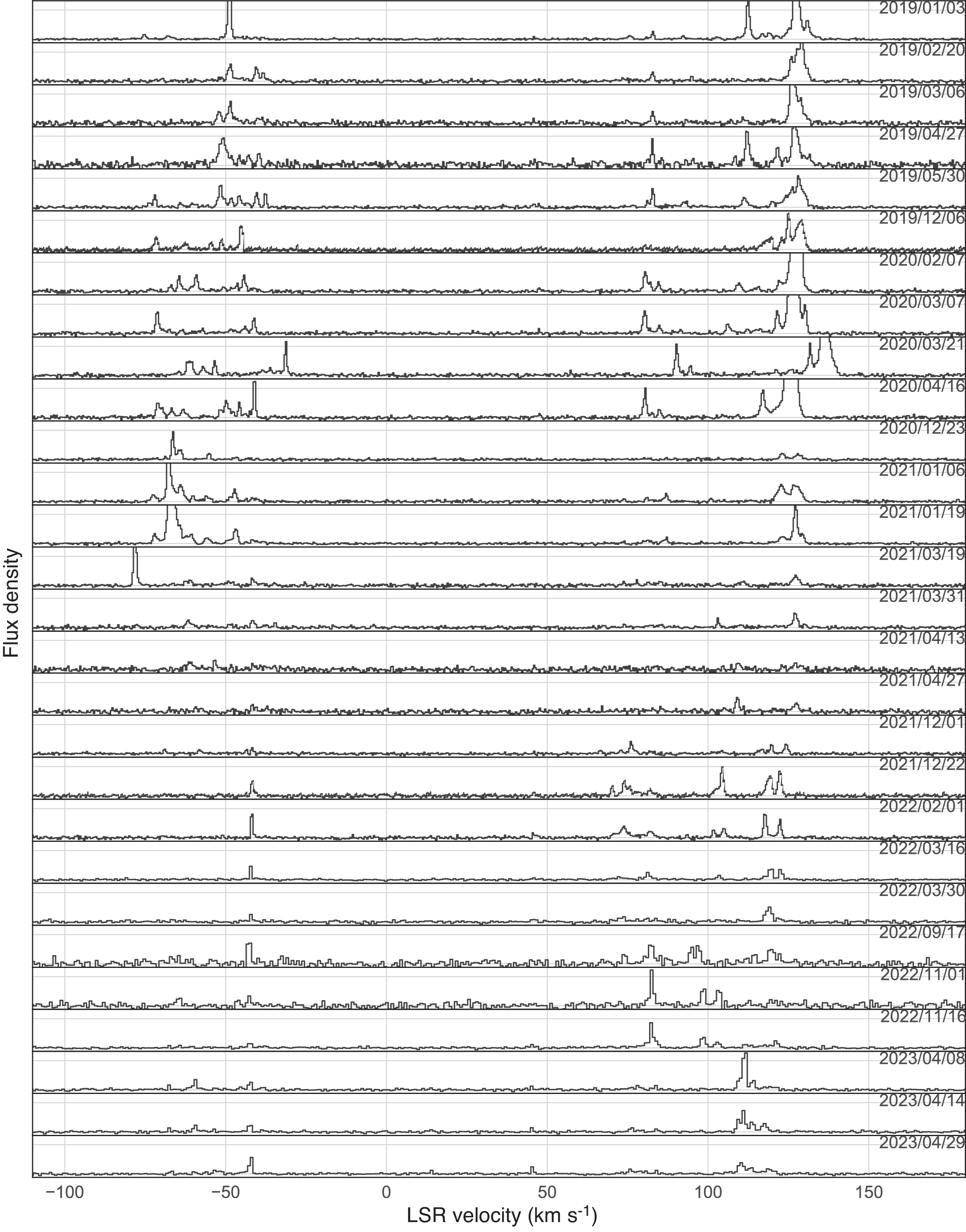} 
\caption{Time series of the H$_2$O maser spectra for IRAS~16552$-$3050. 
The flux density range from $-$0.3~Jy to 3.3~Jy is displayed.}
\label{fig:IRAS16552-3050}
\end{figure*}

Figures \ref{fig:IRAS16552-3050} to \ref{fig:IRAS19190+1102} show the time series of the H$_2$O maser spectra of the WFs. Only the spectral profiles close to the zero-level baselines are displayed to emphasize the emergence of weak spectral features in the high velocity ranges. 
Because the spectra of IRAS~18455$+$0448 and IRAS~18596$+$0315 have exhibited only low velocity components and one of the double-horn components (around $\sim +$115~km~s$^{-1}$), respectively, they are not displayed in this appendix. 

\begin{center}
{\bf These figures are available only in the published PASJ paper.} 
\end{center}

\section{Time series of SiO maser spectra taken for IRAS~16552$-$3050}
\label{sec:spectra-SiO}

Figures \ref{fig:IRAS16552-3050_SiOv1} and \ref{fig:IRAS16552-3050_SiOv2} show the time series of the SiO $v=1$ and $v=2$ $J=1\rightarrow 0$ maser spectra, respectively, for IRAS~16552$-$3050 similarly to the H$_2$O maser spectra (Appendix \ref{sec:spectra}). 

\begin{center}
{\bf These figures are available only in the published PASJ paper.} 
\end{center}

\section{Dynamic spectra of H$_2$O maser spectra taken for the WFs}
\label{sec:dynamic-spectra}

\begin{figure}[ht]
\includegraphics[width=8.3cm]{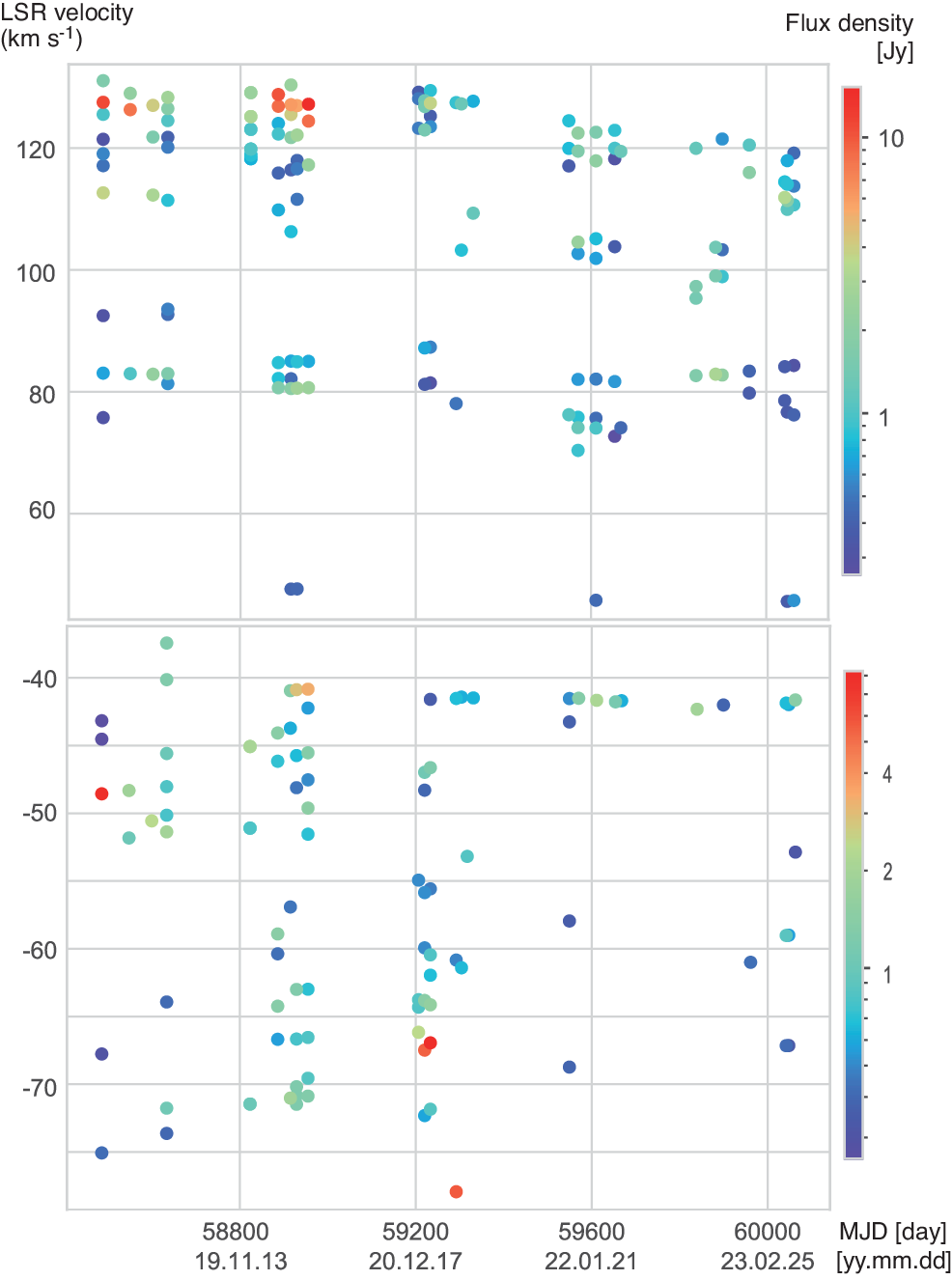}
\caption{Dynamic spectra tracing spectral peaks of H$_2$O masers toward IRAS~16552$-$3050.} 
\label{fig:IRAS16552-3050-DS}
\end{figure}

\begin{figure}[ht]
\includegraphics[width=8.3cm]{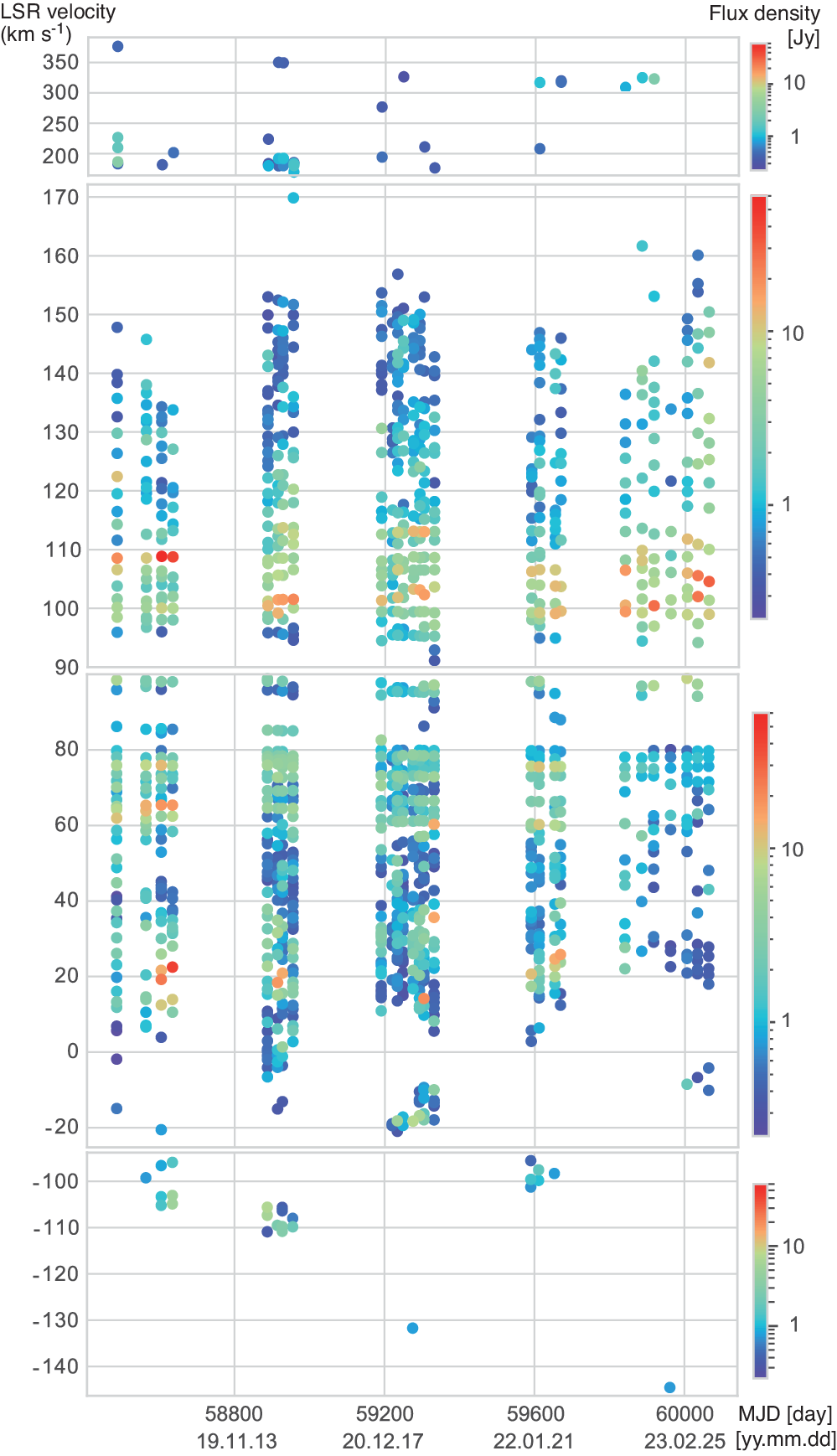}
\caption{Same as figure \ref{fig:W43A-DS} but for IRAS~18043$-$2116.}
\label{fig:IRAS18043-2116-DS}
\end{figure}

\begin{figure}[ht]
\includegraphics[width=8.3cm]{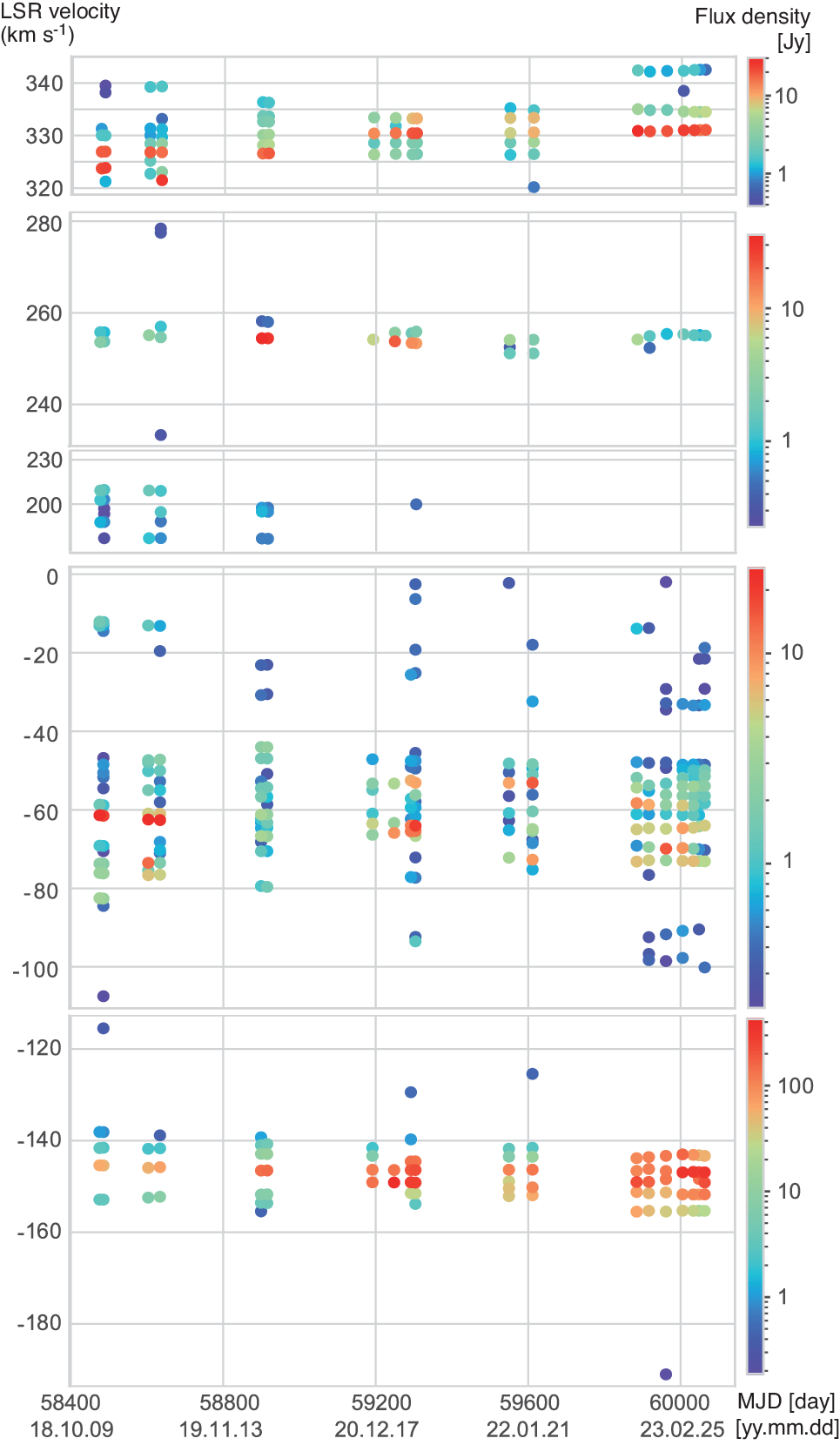}
\caption{Same as figure \ref{fig:W43A-DS} but for  IRAS~18113$-$2503.}
\label{fig:IRAS18113-2503-DS}
\end{figure}

\begin{figure}[ht]
\includegraphics[width=8.3cm]{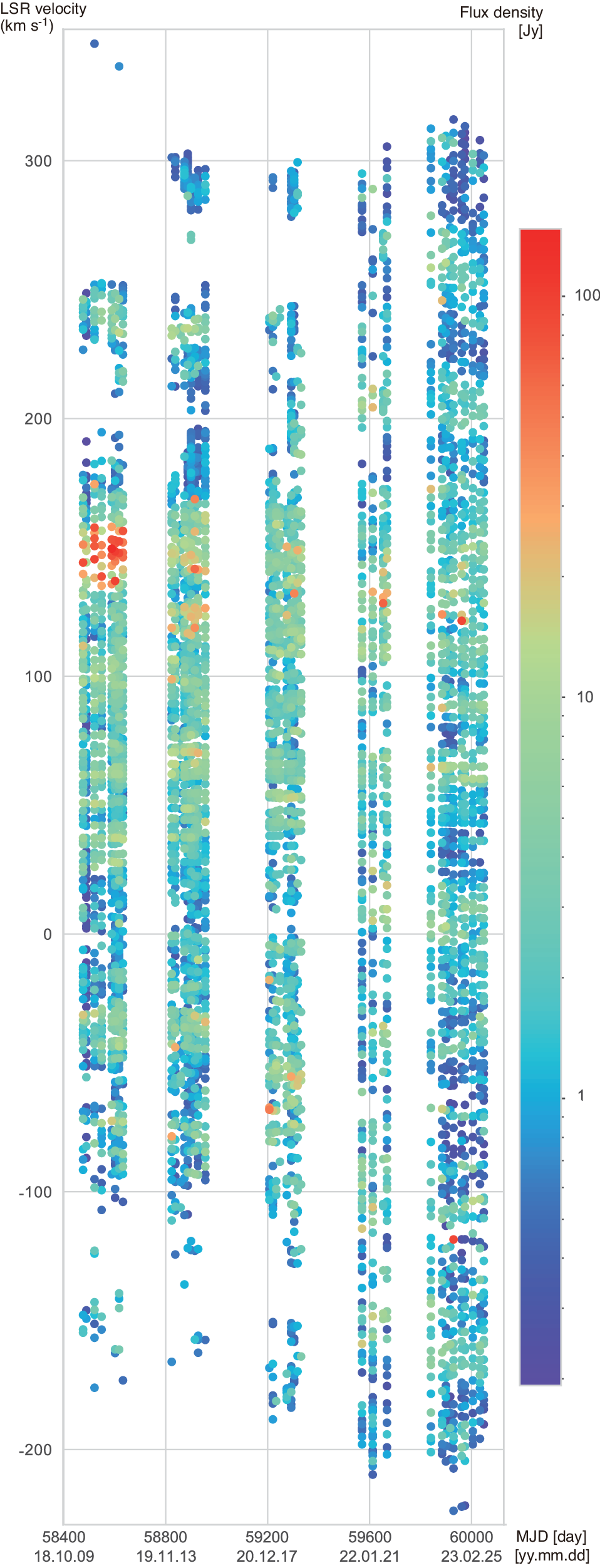}
\caption{Same as figure \ref{fig:W43A-DS} but for IRAS~18286$-$0959.}
\label{fig:IRAS18286-0959-DS}
\end{figure}

\begin{figure}[ht]
\includegraphics[width=8.3cm]{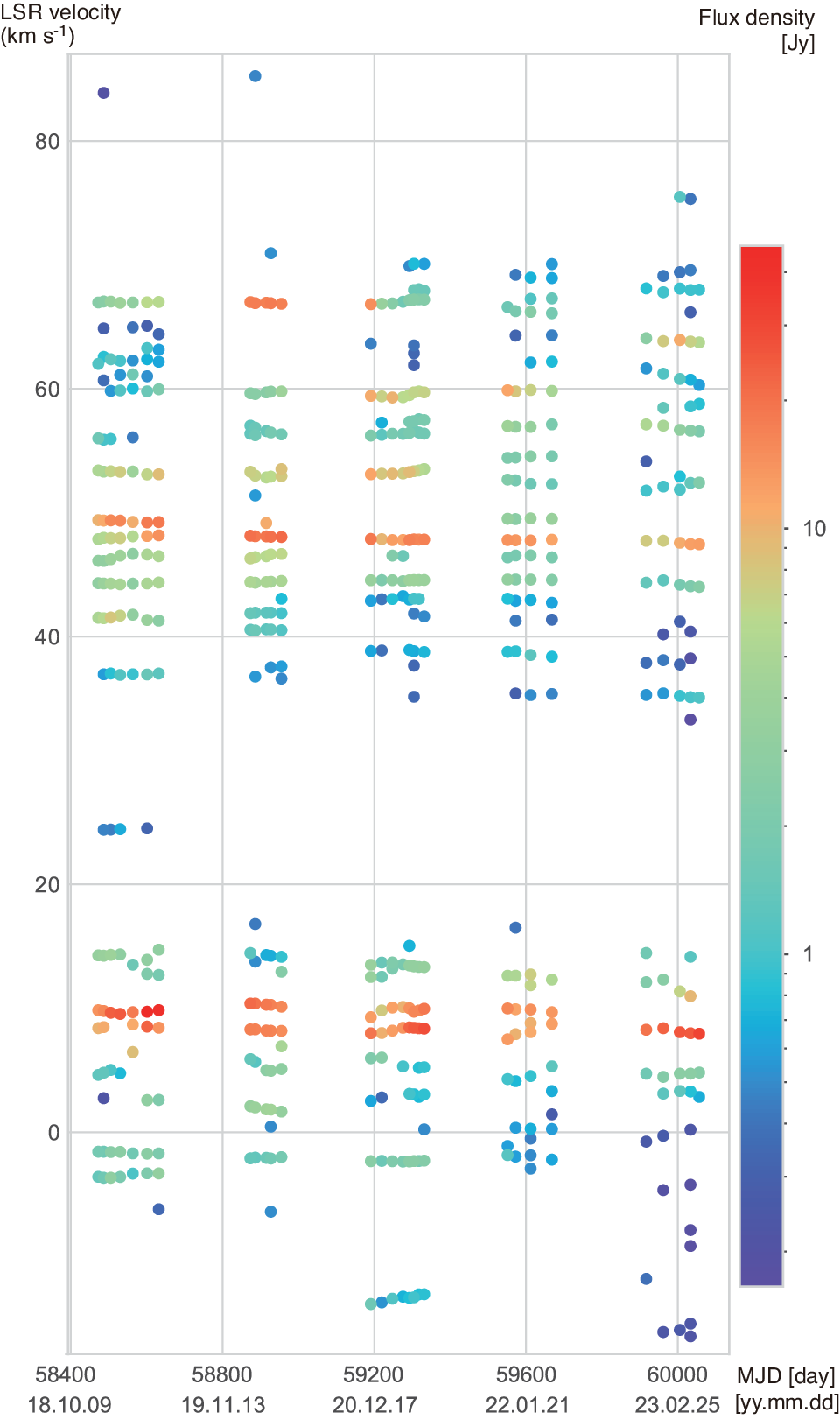}
\caption{Same as figure \ref{fig:W43A-DS} but for IRAS~19190$+$1102.}
\label{fig:IRAS19190+1102-DS} 
\end{figure}

Figures \ref{fig:IRAS16552-3050-DS} to \ref{fig:IRAS19190+1102-DS} show the dynamic spectra of H$_2$O masers tracing spectral peaks for IRAS~16552$-$3050, IRAS~18043$-$2116, IRAS~18113$-$2503, IRAS~18286$-$0959, and IRAS~19190$+$1102, respectively. 




\begin{figure*}[ht]
\caption{Same as figure \ref{fig:IRAS16552-3050} but for IRAS~17291$-$2147. 
The flux density range from $-$0.3~Jy to 1.0~Jy is displayed.}
\label{fig:IRAS17291-2147}
\end{figure*}

\begin{figure*}[ht]
\caption{Same as figure \ref{fig:IRAS16552-3050} but for IRAS~18043$-$2116. 
The flux density range from $-$0.3~Jy to 1.5~Jy is displayed.}
\label{fig:IRAS18043-2116}
\end{figure*}

\begin{figure*}[ht]
\caption{Same as figure \ref{fig:IRAS16552-3050} but for IRAS~18113$-$2503.
The flux density range from $-$0.3~Jy to 3.3~Jy is displayed.}
\label{fig:IRAS18113-2503}
\end{figure*}

\begin{figure*}[ht]
\caption{Same as figure \ref{fig:OH12.8-0.9} but for OH~12.8$-$0.9. 
The flux density range from $-$0.3~Jy to 1.0~Jy is displayed.}
\label{fig:OH12.8-0.9}
\end{figure*}

\begin{figure*}[ht]
\caption{Same as figure \ref{fig:IRAS16552-3050} but for IRAS~18286$-$0959.
The flux density range from $-$0.3~Jy to 3.3~Jy is displayed.}
\label{fig:IRAS18286-0959}
\end{figure*}

\begin{figure*}[ht]
\caption{Same as figure \ref{fig:IRAS16552-3050} but for W~43A. 
The flux density range from $-$0.3~Jy to 3.3~Jy is displayed.}
\label{fig:W43A}
\end{figure*}

\begin{figure*}[ht]
\caption{Same as figure \ref{fig:IRAS16552-3050} but for IRAS~18460$-$0151. 
The flux density range from $-$0.3~Jy to 3.3~Jy is displayed.}
\label{fig:IRAS18460-0151} 
\end{figure*}

\begin{figure*}[ht]
\caption{Same as figure \ref{fig:IRAS16552-3050} but for IRAS~19134$+$2131. 
The flux density range from $-$0.3~Jy to 3.3~Jy is displayed.}
\label{fig:IRAS19134+2131}
\end{figure*}

\begin{figure*}[ht]
\caption{Same as figure \ref{fig:IRAS16552-3050} but for IRAS~19190$+$1102. 
The flux density range from $-$0.3~Jy to 3.3~Jy is displayed.}
\label{fig:IRAS19190+1102}
\end{figure*}

\begin{figure*}[ht]
\caption{Time series of the SiO $v=1$ $J=1\rightarrow 0$ maser spectra for IRAS~16552$-$3050. 
The flux density range from $-$0.4~Jy to 1.0~Jy is displayed.}
\label{fig:IRAS16552-3050_SiOv1}
\end{figure*}

\begin{figure*}[ht]
\caption{Same as figure \ref{fig:IRAS16552-3050_SiOv1} 
but for the SiO $v=2$ $J=1\rightarrow 0$ maser spectra for IRAS~16552$-$3050.}
\label{fig:IRAS16552-3050_SiOv2}
\end{figure*}

\end{document}